\documentstyle[eqsecnum,aps]{revtex}
\def\laq{\ \raise 0.4ex\hbox{$<$}\kern -0.8em\lower 0.62
ex\hbox{$\sim$}\ }
\def\gaq{\ \raise 0.4ex\hbox{$>$}\kern -0.7em\lower 0.62
ex\hbox{$\sim$}\ }                            
\def\half{\hbox{\magstep{-1}$\frac{1}{2}$}}
\def\quarter{\hbox{\magstep{-1}$\frac{1}{4}$}}
\def\NPB{{\em Nucl. Phys.} B}
\def\PLB{{\em Phys. Lett.}  B}
\def\PRL{{\em Phys. Rev. Lett.}}
\def\PRD{{\em Phys. Rev.} D}
\def\MPL{{\em Mod. Phys. Lett.}  A}
\begin{document}
\title{Particle production in string cosmology models}

\author{Ram Brustein} 
\address{Department of Physics,
Ben-Gurion University,
Beer-Sheva 84105, Israel, and \\
Theory Division, CERN, CH-1211, Geneva 23, Switzerland\\
email: ramyb@bgumail.bgu.ac.il}

\author{Merav Hadad} 
\address{School of Physics and Astronomy,
Beverly and Raymond Sackler Faculty of Exact Sciences,\\
Tel Aviv University, Tel Aviv 69978, Israel\\
email: meravv@post.tau.ac.il}

\maketitle
\begin{abstract}

We compute spectra of  particles produced during a dilaton-driven kinetic
inflation phase within string cosmology models.  The resulting spectra depend
on the parameters of the model and on the type of particle and are quite
varied, some increasing and some decreasing with frequency. We use an
approximation scheme in which all spectra  can be expressed in a nice
symmetric form, perhaps hinting at a deeper symmetry of the underlying
physics.  Our results may serve as a starting point for detailed studies of
relic abundances, dark matter candidates, and possible sources of large scale
anisotropy.
\end{abstract}
\centerline{Preprint Numbers: BGU-PH-97/12, TAUP-2448-97}

\section{INTRODUCTION}

An inflationary scenario \cite{gv1,gv2} (the so called ``pre-big-bang"    
scenario) postulates that the evolution of the Universe starts from a state of
very small curvature and coupling and then undergoes a long phase of
dilaton-driven kinetic inflation, in which the curvature and coupling grow, 
and then at some later time joins smoothly standard  Friedman-Robertson-Walker
(FRW) radiation dominated cosmological evolution, thus giving rise to a
singularity free inflationary cosmology. Since the graceful exit transition
from the dilaton-driven phase to the decelerated FRW evolution cannot occur
while curvatures and coupling are small \cite{bv,kmo}, an intermediate high
curvature ``string phase" \cite{bggv}, in effect replacing the big bang, has
to separate the early inflationary phase from the late FRW phase. The required
initial conditions of the Universe  in this scenario were discussed
\cite{inhom1,inhom2,inhom3}, and  the graceful exit transition is now better
understood \cite{gmv,bm}. We assume that the appropriate initial conditions
were chosen such that a long dilaton-driven inflationary phase was indeed part
of the evolution and that a later standard FRW cosmology followed, and wish to
study possible consequences. 

We study production of a variety of particle types during the inflationary
dilaton-driven phase. So far, most of the  produced particle spectra were
found to rise sharply with frequency. Therefore, in these spectra, most of the
power is concentrated at higher frequencies, a property with interesting
consequences \cite{bggv,gravitons,photons,dilatons}, but also some
disadvantages. In particular, at very large wavelengths there is almost no
power, making these inhomogeneity perturbations an unlikely source for large
scale anisotropy observed in the cosmic microwave background and required as
seeds for structure formation. The standard explanation given for the generic
spectral frequency dependence is that since the curvature increases sharply,
particle production also increases and hence the resulting spectrum. Certain
axionic perturbations were so far the only exception found to this tendency
\cite{axions}. 
We find that although sharply rising spectra are indeed common,  flatter
spectra or even decreasing spectra are just as likely. We find that the slope
of the spectrum depends on the spin of the particle and on the type of dilaton
prefactor and whether it is massless or massive, revealing a  rich range  of
spectral shapes, of which many more deserve further individual attention.

The model of background evolution we adopt in this paper is a simplified model.
The evolution of the Universe is divided into four distinct  phases, the first
phase is a long dilaton-driven inflationary phase. We assume that the
background solution is the simplest solution of the string-dilaton-gravity
equations of motion in 4 dimensions, the so called  (+) branch vacuum. The
second phase is a high-curvature string phase of otherwise unknown properties,
followed by ordinary FRW radiation dominated (RD) evolution and then a standard
FRW matter dominated (MD) evolution, both with a fixed dilaton.  We assume that
curvature stays high during the string phase, in a sense that we define better
later, but we do not assume any specific form of background evolution.
Obviously, by doing that we give up on the possibility of obtaining any
information about the produced particles during the string phase. We postpone
such analysis until some better ideas and more reliable methods for handling
the string phase are at hand (see, however, \cite{bggv,stph}). We do not lose the
ability to compute the spectra produced during the dilaton-driven phase, as we
explain later. We parametrized our ignorance about the string phase background,
as in \cite{bggv}, by the ratios of the scale factor and the string coupling, 
at  moments in (conformal) time marking the beginning and end of the string
phase $z_S=a_1/a_S$ and $g_1/g_S$. It turns out that in the generalized setting we use,
these two parameters are still sufficient to parametrize all spectra.

Particles get produced during the dilaton-driven phase by the standard
mechanism of amplification of vacuum fluctuations \cite{birrdav,pert}.
Deviations from homogeneity and isotropy are generated by quantum 
fluctuations around the homogeneous and isotropic background and then
amplified by the accelerated expansion of the Universe.  In practice, we
compute particle production in the standard way, by solving  linearized
perturbation equations with vacuum fluctuations boundary conditions.  The
general solution of the perturbation equation is a linear combination of two
modes, one which is approximately constant  for wavelengths large compared
with the curvature of the Universe (outside the horizon),  and one which is
generically time dependent outside the horizon. We understand the appearance
of a constant mode as the freezing of the perturbation amplitude, since local
physics is no longer active on such scales. The  existence of the time
dependent mode can be most easily understood in terms of a constant mode of
the conjugate momentum of the perturbation \cite{sbarduality}. The amplitude
of the conjugate momentum also freezes outside the horizon,  since local
physics is no longer active on such scales. This forces the time derivative of
the perturbation to be non-vanishing,  leading to a ``kinematical" time
dependence of the perturbation amplitude. The analysis of perturbation
equations and their the solutions  in the Hamiltonian formalism was first
suggested in \cite{sbarduality}, as a way of understanding certain
symmetries of the spectra of gravitons and photons.

In addition to the dilaton and metric, string theory contains many other fields
that have trivial expectation values in our model of background evolution and
do not affect the classical solutions, but they do fluctuate. We are interested
in ``low-mass" fields, which are either massless, or have masses much below
the string scale, such as moduli, gauge bosons, and their superpartners.   In
realistic models the mass of different fields may depend in a complicated way
on the dilaton or other moduli. We will take the mass as an additional
parameter and assume that it vanishes in the dilaton-driven phase and takes a
constant value, much smaller than the string mass $M_s=M_p g_1$ ($M_p$ is the
Planck mass), from the start of the RD era and on. This assumption could be
relaxed without substantial changes in results. For simplicity, we assume that
the produced particles are stable particles, decoupled from the radiation
thermal bath. Adding the effects of interactions with the thermal bath and
decay is rather straightforward and has to be done on a case by case basis.

For a large class of particles, it is enough to specify a particle type with
two parameters, $m$ and $l$, specifying the interaction of the field with the
background metric and dilaton. We explain the appearance of these parameters
in more detail later, but for now we give the following list for  some
particular cases. For the dilaton or graviton $m=1$ and $l=-1$, for the model
independent axion $m=l=1$, for perturbative gauge fields such as photons,
$m=0$ and $l=-1$, while for some non-perturbative gauge fields $m=0$ and
$l=0$, Ramond-Ramond axions and non-perturbative scalars have $m=1$ and $l=0$.
In general, fields with more tensor indices tend to have smaller (more
negative) $m$'s.  We do not know at the moment a general rule for allowed
values of $m$ and $l$, but it is obvious that many combinations are possible. 

Possible benefits of our results are expected to come from several different
aspects. First, from the  different dependence on the fundamental parameters
$g_S$, $z_S$ of decreasing as well  as increasing spectra. This leads to
different constraints on the allowed range of parameters, allowing to narrow
their range and give better prediction for the spectrum of relic gravitational
radiation background. Flatter spectra allow the possibility of producing large
scale inhomogeneities with enough power to explain observed cosmic microwave
background anisotropies. Another exciting possibility is that some of the
weakly interacting particles will constitute  the dark matter
in the Universe. Our results provide a starting point for actual calculations
regarding these issues and we hope to study them in the future. 

Until now, the spectra that were computed were the following. Production of
gravitons, for which the full dependence on parameters, effects of late time
evolution and observability were taken into account \cite{bggv,gravitons}.
Production of dilatons was partially studied \cite{dilatons}, not taking into
account the effects of an intermediate string phase. Photon production was
studied \cite{photons}, taking into account dependence on parameters and late
time evolution. Certain axionic perturbations were studied without
taking into account the effects of an intermediate string phase and the
effects of mass \cite{axions}. In \cite{bmvu} some additional fields were
studied.

Because this work is the first attempt to make a comprehensive  survey of
particle production in string cosmology models, in the hope of discovering some
interesting phenomena, as we indeed do, we adopt a modest, minimalist
approach.  Our main goal in this paper is to show that spectra of produced
particles come in different shapes, and we feel that this point is better
demonstrated with spectra than can be reliably computed. We use approximations
that we judge not to hide our main results and are accurate enough for our
purposes.  There are, however, many obvious improvements,
which could be implemented if necessary. We would like to issue a warning about
numerical coefficients, which should not be taken too seriously, in view of the
approximations we used. In the same spirit we make no attempt to obtain  bounds
from constraints on the total energy in fluctuations, or to require that the
total number of particles leads to the observed radiation (as in \cite{bmvu}),
or to impose any other additional requirements on the spectra. We believe that
this should be done for each class of particles separately, with additional
input in the form of assumptions about stability, interactions and dependence
on late stages of the evolution. We are completing a study of relic abundance 
axions \cite{bh}, and plan to study some additional interesting cases. 

The plan of the paper is as follows. In section 2 we describe in more detail
the model of background evolution, the perturbation equations, and their
boundary conditions. In section 3 we describe  our method of solving the
perturbation equations, and solve them. The paper is quite technical and we
have made a special effort to summarize our results in a self-contained form in
section 4. The last section contains our conclusions and some preliminary
consequences.

\section{Background evolution and particle production}

In this section we describe the model of background evolution, the
perturbation equations obeyed by the different fields, and our method of
solving these equations.

\subsection{The model of background evolution}

The model of background evolution we adopt in this paper is a simplified
model. The evolution of the Universe is divided into four distinct 
phases, the first phase is a long dilaton-driven
inflationary phase, the second phase is a high-curvature string phase of
otherwise unknown properties, followed by ordinary FRW RD
evolution and then a standard FRW MD evolution. 
We assume throughout an isotropic and homogeneous
 four dimensional flat Universe, described by a FRW metric with the 
line element 
$
ds^2=a^2(\eta )\left( d\eta ^2-\delta _{ij}dx^idx^j\right),  
$
where $\eta$ is conformal time and $a(\eta )$ is the scale factor. The dilaton
$\phi$ is 
time dependent, $\phi=\phi(\eta)$. All other fields that we will 
be interested in are assumed to have a trivial VEVs. 
Therefore, to specify the phases in a
concrete way, we need to specify  two functions, $a(\eta)$ and
$\phi(\eta)$ and the conformal time boundaries between the phases, which we
proceed to do.
\begin{itemize} 
\item
The dilaton-driven inflationary phase lasts while  
$\eta <\eta _s$,
and while it lasts the scale factor and dilaton are given by the solution
of the lowest order string-dilaton-gravity equations of motion, the
so-called $(+)$ branch vacuum,
\begin{equation}
a=a_s\left( \frac \eta {\eta _s}\right) ^{(1-\sqrt{3})/2}, \hspace{.5in}
e^\phi =e^{\phi _s}\left( \frac \eta {\eta _s}\right) ^{-\sqrt{3}}_{.} 
\label{ddphi}
\end{equation}
where $a_s=a(\eta_s)$ and $\phi_s=\phi(\eta_s)$. Both curvature and coupling
are growing in this phase, which is expected to last until the time $\eta _s$
when the curvature reaches the string scale  and the background solution
starts to deviate substantially from the lowest order solution. For recent
ideas about how this may come about see \cite{gmv,bm}.

\item

The string phase lasts  while  $\eta _s<\eta <\eta _1$. We assume that
curvature stays high during the string phase. As in \cite{peak}, we assume
that  the string phase ends when $H(\eta_1)/a(\eta_1)\simeq M_s$. We therefore
implicitly assume that the string coupling at the end of the string phase is
higher than it's value at the beginning of the string phase and that it does
not decrease much during the string phase. But we do not assume any specific
form of background evolution. Obviously, by doing that we give up on the
possibility of obtaining any information about the produced particles during
the string phase. We postpone such analysis until some better ideas and more
reliable methods for handling the string phase are at hand (see, however,
\cite{stph}). We do not lose the ability to compute the spectra produced during
the dilaton-driven phase, as we explain later. We parametrized our ignorance
about the string phase background, as in \cite{bggv}, by the ratios of the
scale factor and the string coupling $g(\eta)=e^{\phi(\eta)/2}$,  at the
beginning and end of the string phase $z_S=a_1/a_S$ and $g_1/g_S$, where
$g_1=e^{\phi(\eta_1)/2}$ and  $g_S=e^{\phi(\eta_S)/2}$.  It turns out that also
in the generalized setting we use, these two parameters are enough. We take the
parameters to be in a range we consider reasonable, for example, $z_S$ could be
in the range $1<z_S<e^{45}\sim 10^{20}$, to allow a large part of the observed
Universe to originate in dilaton-driven phase, and $g_1/g_S>1$. Note
that unlike \cite{bggv} or \cite{stph} we do not assume a specific background
during the string phase. The time $\eta_1$ marks the start of the RD era as we
explain below. Some other useful quantities that we will need are $\omega_1$,
the frequency today, corresponding to the end of the string phase, estimated in
\cite{peak}, and the frequency $\omega_s= \omega_1/z_S$, the frequency today
corresponding to the end of the dilaton-driven phase.

Our definition of the string phase
encompasses the whole duration in between the end of the dilaton-driven phase
and the onset of RD phase, which may be quite complicated as in \cite{bm}.

\item The FRW RD phase is assumed to follow the string  phase and last while
$\eta _{1}<\eta <\eta _{eq}$, \begin{equation} a=a_1\left( \frac \eta {\eta
_1}\right),\hspace{.5in}  e^\phi =e^{\phi _1}=const. \end{equation} where $\eta
_{eq}$ is the time of matter-radiation equality, $a_1=a(\eta_1)$ and
$\phi_1=\phi(\eta_1)$ . Note that the dilaton is taken to be strictly constant,
frozen at its value at $\eta=\eta_1$.  \item The FRW MD phase is assumed to 
last while $\eta _{eq}<\eta <\eta _{0}$ \begin{equation} a=a_{eq}\left( \frac
\eta {\eta _{eq}}\right) ^2,\hspace{.5in} e^\phi =e^{\phi _1}=const.
\end{equation} where $\eta_0$ is today's conformal time and
$a_{eq}=a(\eta_{eq})$.  Note that the constant value of the dilaton is further
assumed to be that of today. Some possible consequences of late time evolution
of the dilaton were discussed in \cite{giovani}. 
\end{itemize} 
For more accurate results, we could have used a more sophisticated description
of the RD and MD phases using entropy conservation, effective number of degrees
of freedom and temperature, as was done in \cite{peak}. 

We presented our model in the string frame, using string frame conformal
time. Another favorite choice is the Einstein frame, in which the
gravitational action takes the standard Einstein form. We continue to use the
string frame exclusively throughout the paper. General arguments \cite{gv3},
and many explicit examples have shown that physical
quantities such as energy density or number density can be computed in any
frame.

\subsection{The perturbation equation}

Deviations from homogeneity and isotropy are generated by quantum  fluctuations
around the homogeneous and isotropic background and then amplified by the
accelerated expansion of the Universe. In addition to the dilaton and metric,
string models contain many other fields that have trivial expectation values in
our model of background evolution and do not affect the classical solutions, but
they do fluctuate. We are interested in ``low-mass" fields, which are either
massless, or have masses much below the string scale, such as moduli, gauge
bosons, and their superpartners. 

The action for each field's perturbation is obtained by expanding the 4
dimensional effective action of strings, which generically, for a tensor field
of rank $N$, has the form  
$
{\small\frac{1}{2}} \int d^4 x \sqrt{-g}
e^{l\phi}{\cal L}^2 (T_{\mu_1,...,\mu_N})
$, where the parameter $l$ in the
dilaton prefactor is determined by the type of field. The second order
covariant Lagrangian ${\cal L}^2$ contains a kinetic term and possibly also a
mass term. Setting the dilaton and metric at their background 
values, $g_{\mu\nu}(\eta)=a^2(\eta )\eta_{\mu\nu}$,  $\phi(\eta)$ results in a
 quadratic action for each physical component of the perturbation,
which we denote by $\psi$,  
\begin{equation} 
A={\small\frac{1}{2}} \int d\eta\ d^3x\ a^{2m}e^{l\phi }
\left( \psi ^{\prime 2}- (\nabla \psi )^2-M^2 a^2 \psi ^2\right)    
\label{pertact}  
\end{equation} 
where $'$ denotes $\partial/\partial\eta$, $M$ is the mass of the perturbed
field and $m$ and $l$ depend on the type of field.  We ignore complication
associated with gauge invariance and work directly with physical components.
For example, the background metric has 10 components of them 8 are
non-physical, and each of the two physical components has an effective action
of the form (\ref{pertact}).

As already mentioned, we are interested in particles for which the mass $M$
is either vanishing, or much smaller than $M_s$. In realistic models the mass
may depend in a complicated way on the dilaton or other moduli. We will take
$M$ as an additional parameter and assume that it vanishes in the
dilaton-driven phase and takes a constant value, much smaller than $M_s$, from
the start of the RD era and on. The parameters $l$ and $m$ take the following
values in some particular cases. For the dilaton or graviton $m=1$ and
$l=-1$, for the model independent axion $m=l=1$, for perturbative gauge
fields such as photons, $m=0$ and $l=-1$, while for some non-perturbative
gauge fields $m=0$ and $l=0$, Ramond-Ramond axions and non-perturbative
scalars have $m=1$ and $l=0$. In general, fields with more tensor indices
tend to have smaller (more negative) $m$'s. For example, the physical
components of a 2-index tensor, whose field-strength is a 3-index tensor
could have $m=-1$. We do not know at the moment a general rule for allowed
values of $m$ and $l$, but it is obvious that many combinations are
possible.

The linearized equation of motion, satisfied by the field perturbation $\psi$,
derived from the action (\ref{pertact}) is the following, 
\begin{equation} 
\left( \triangle +M^2\right) \psi +\left( l+( m-1)\left( 1-1/\sqrt{3}
 \right) \right) a^{-2}\psi ^{\prime }\phi ^{\prime }=0 
\label{perteq1} 
\end{equation}

where $\triangle \psi$ is the covariant Laplacian and we have used the fact
that for our particular case $e^{\phi(\eta)} =a(\eta)^{3+\sqrt{3 }}$. 
In Fourier space
$\triangle \psi =
a^{-2}\left[ \psi_k ^{\prime \prime }+2H\psi_k ^{\prime}+k^2\psi_k \right] $ 
and therefore the perturbation equation (\ref{perteq1}) is the following
\begin{equation}
\psi_k ^{\prime \prime }+
\left[ 2H+\left( l+\left( m-1\right) 
\left( 1-1/\sqrt{3}\right) \right) 
\phi ^{\prime }\right] \psi_k ^{\prime }
+\left(k^2+M^2a^2\right) \psi_k =0.  
\label{perteq2}
\end{equation}
Using the transformation $\chi =S(\eta)\ \psi$ where $S(\eta)=a(\eta)^m
g(\eta)^l$ and $g(\phi)=e^{\phi /2}$, eq.(\ref{perteq2}) can be
simplified,
\begin{equation}
\chi_k ^{\prime \prime }+
\left( k^2+M^2a^2-\frac{S^{\prime \prime }}S\right) \chi_k =0.  
\label{perteq3}
\end{equation}
The function $S(\eta)$ was first introduced in \cite{sbarduality}.

\subsection{Approximate forms and solutions of the perturbation equation}

Before trying to actually solve eq.(\ref{perteq3}) it is useful to look at
approximate forms of the equation and their solutions in different physical
situations. 

A few special moments exist in the life of a perturbation, first,
$\eta\rightarrow -\infty$ when it is essentially a free field perturbation in
flat space-time. Then, the time $\eta_{ex}$ when the rate of expansion $H$
becomes too fast for a given perturbation and it ``exits" the horizon and its
amplitude freezes (see below). The perturbation remains ``outside the horizon"
until the ``reentry" time $\eta_{re}$ when the expansion rate $H$ is again slow
enough and the perturbation defreezes. We are interested in perturbations that
exit during the dilaton-driven phase $\eta_{ex}<\eta_s$, and reenter during
RD,  $\eta_{re}>\eta_1$. We do assume that all perturbations stay  outside the
horizon during the string phase, using in an essential way that the curvature
during the string phase stays high.

In the range $\eta <\eta _{ex}$,  $k^2> S^{\prime\prime}/S\propto H^2$,   in
the range $\eta _{ex}<\eta <\eta _{re}$,  
$k^2+M^2a^2<S^{\prime\prime}/S\propto H^2$ and in the range $\eta>\eta_{re}$,  
$k^2+M^2a^2> S^{\prime\prime }/S\propto H^2$.  Since we are interested in
fields whose mass gets generated after the onset of RD and then have small
masses,  we observe that in the whole  range $-\infty <\eta <\eta _{re}$ the
mass term in the perturbation equation may be ignored. In effect,
in this range of time, we may use the approximate
form
\begin{equation} 
\chi_k ^{\prime \prime }+ \left( k^2-
\frac{S^{\prime \prime}}S\right) \chi_k =0. 
\end{equation}

The general solution of the perturbation equation is a linear combination of
two modes, one which is approximately constant outside the horizon and one
which is generically time dependent outside the horizon. We understand the
appearance of a constant mode as the freezing of the perturbation amplitude,
since local physics is no longer active on such scales. The  existence of the
time dependent mode can be most easily understood in terms of a constant mode
of the conjugate momentum of $\psi$, $\Pi=S^2(\eta)\psi'$. The amplitude of
the conjugate momentum also freezes outside the horizon,  since local
physics is no longer active on such scales. This forces $S^2(\eta)\psi'$ to be
approximately constant leading to a ``kinematical" time dependence of $\psi$.
The perturbation for $\Pi_k$ is simplified if
rewritten for ${\chi^\pi}_{k} =S^{-1}\Pi_k$ 
\begin{equation}
{\chi^\pi}_{k}  ^{\prime \prime }+
\left( k^2-\frac{(S^{-1})^{\prime \prime }}{ 
S^{-1}}\right) {\chi^\pi}_{k} =0
\end{equation}
and it's solutions are Bessel functions, as for $\chi_k$. 

In the RD Universe (MD Universe) $a=a_1\eta
/\eta _1$ ($a=a_1\eta _{eq}/\eta _1\left( \eta /\eta _{eq}\right) ^2$), thus
even if the $S^{\prime \prime }/S$ is dominant at the beginning of the
RD phase, then at later times $M^2a^2+k^2$ becomes
dominant  and therefore in the range $\eta >\eta _{re}$ the perturbation
equation may be approximated by
\begin{equation}
\chi_k ^{\prime \prime }+\left( k^2+M^2a^2\right) \chi_k =0.
\label{pertsolmrd}
\end{equation}

The number density and energy density of the produced particles can be easily
read off from the solutions of eq.(\ref{pertsolmrd}) in a standard way
\cite{birrdav}.

We solve the perturbation equation using the following general method. We
solve the early time equation with boundary conditions of normalized vacuum
fluctuations. We assume that the constant mode of $\psi$ and the constant
mode of $\Pi$ remain constant also while outside the horizon during the
string phase, thus bridging the gap of unknown background evolution during
the string phase. We then match the early time solutions to the late time
solutions at $\eta=\eta_1$ when both the early time and late time equation
are expected to be good approximations. For massive particles, an additional
matching step is performed at $\eta=\eta_{re}$. In all known cases our method 
reproduces results obtained by assuming specific background evolution during
the string phase. 

\section{Solutions of the perturbation equation}

In this section we solve the perturbation equation, using the general
procedure outlined in the previous section and obtain the necessary
ingredients to assemble the spectra of produced particles.

\subsection{Early time solutions}

As explained previously, the approximate perturbation equation for 
early times
is given by
\begin{equation}
\chi_k ^{\prime \prime }+
\left( k^2-\frac{S^{\prime \prime }}S\right) \chi_k =0.
\label{earlyperteq}
\end{equation}
In the range $\eta <\eta _s,$ $S(\eta)=a_s^me^{l\phi _s/2}\left( \eta
/\eta _s\right) ^{1/2-n_s}$ where 
$n_s=\left( \sqrt{3}(l+m)+1-m\right) /2,$
and in the range $\eta >\eta _1,$ 
$S(\eta)=a_1^me^{l\phi_1/2}\left( \eta /\eta_1\right)^{1/2-n_1}$ 
where  $n_1=\half-m.$
Therefore, $S^{\prime \prime }/S=(n^2-\quarter)\eta ^{-2},$ 
where $n$ stands for
either $n_s$ or $n_1$, and the perturbation
equation (\ref{earlyperteq}) takes the following simple form
\begin{equation}
\chi_k ^{\prime \prime }+
\left( k^2-(n^2-\quarter)\eta ^{-2}\right) \chi_k =0.
\label{earlyperteq1}
\end{equation}
The solutions of this equation are  Bessel functions. For the special values
$n_s=\pm \half$ (corresponding for example to $m=0$, $l=0$),
 or $n_1=\pm \half$ (corresponding to $m=0,1$),
 the equation simplifies and the solutions are simple exponentials. It is
 nevertheless possible to continue to use the general setup and substitute
 these special values at the end.

Imposing that
 for $k\eta \longrightarrow -\infty $ the solution corresponds to vacuum
 fluctuations 
 $\frac{1}{\sqrt{2k}}e^{-ik\eta }$, we obtain for $\eta <\eta _s$ 
\begin{equation}
\chi_k(\eta) =
\frac{\sqrt{\pi }}2\sqrt{\eta }J_n(k\eta )+
\frac{\sqrt{\pi }}2\sqrt{\eta }Y_n(k\eta ),  
\label{earlysol}
\end{equation}
while for for $\eta >\eta _1$ we do not have any restrictions on the
solution,
\begin{equation}
\chi_k =A\sqrt{\eta }J_n(k\eta )+B\sqrt{\eta }Y_n(k\eta ).
\label{latesol}
\end{equation}
For $\eta <\eta _s$, for perturbations outside the horizon, 
$|k\eta|\laq 1$, $\chi_k$
can be approximated by
\begin{equation}
\chi_k =\frac{\sqrt{\pi }}{2}
\frac{ \left( 1-i\cot \left( n_s\pi \right)\right)}
{\Gamma\left( 1+n_s\right)}
\sqrt{\eta }\left( \frac{k\eta }{2}\right)^{n_s} 
\left[ 1-\frac{(k\eta )^2}{2(2+2n_s)}\right]   
-i\frac{\sqrt{\pi }}{2}
\frac{ \csc \left( n_s\pi \right)}{\Gamma\left( 1-n_s\right)}
\sqrt{\eta }\left( \frac{k\eta }2\right)^{-n_s} 
\left[ 1-\frac{(k\eta )^2}{2(2-2n_s)}\right],
\end{equation}
while for $\eta >\eta _1$ and  $k\eta\laq 1$,  $\chi_k$
can be approximated by
\begin{equation}
\chi_k = A
\frac{ \left(1-i\cot \left( n_1\pi\right) \right)} {\Gamma\left( 1+n_1\right)}
\sqrt{\eta }\left( \frac{k\eta }2\right)^{n_1}
\left[1-\frac{(k\eta )^2}{2(2+2n_1)}\right] -
i B \frac{ \csc \left( n_1\pi \right)}{\Gamma\left( 1-n_1\right)}
\sqrt{\eta }\left( \frac{k\eta }2\right)^{-n_1}
 \left[ 1-\frac{(k\eta )^2 
}{2(2-2n_1)}\right]. 
\end{equation}
Therefore, in the range $\eta<\eta _s$ the solution is approximately given by
\begin{eqnarray}
\psi^{in}_k &=& \frac{2^{-1-n_s}}{\sqrt{\pi k_s}}\Gamma \left( -n_s\right)
a_s^{-m}e^{-l\phi _s/2}\left( k_sk\right) ^{n_s}\eta ^{2n_s}\left[ 1-\frac{ 
(k\eta )^2}{2(2+2n_s)}\right] \nonumber \\ &+&
\frac{2^{n_s-1}}{\sqrt{\pi k_s}}\Gamma \left( n_s\right)
a_s^{-m}e^{-l\phi _s/2}\left( \frac k{k_s}\right) ^{-n_s}\left[ 1-\frac{ 
(k\eta )^2}{2(2-2n_s)}\right], 
\label{insol}
\end{eqnarray}
where we have used the relations 
$\Gamma (n)\Gamma (1-n)=\pi /\sin (n\pi)$ 
and $\Gamma(-n)\Gamma (1+n)=\pi \left| 1-i\cot (\pi n)\right|$, 
and defined $k_s\equiv 1/|\eta _s|$. In the range $\eta>\eta _1$ we obtain
\begin{eqnarray}
\psi_k^{out}&=& A\frac{1}{2\pi \sqrt{k_1}}2^{-n_1}\Gamma \left(
-n_1\right) a_1^{-m}e^{-l\phi _1/2}\left( k_1k\right) ^{n_1}\eta
^{2n_1}\left[ 1-\frac{(k\eta )^2}{2(2+2n_1)}\right] \nonumber \\ &+&
B \frac{1}{\pi \sqrt{k_1}}2^{n_1}\Gamma \left( n_1\right)
a_1^{-m}e^{-l\phi _1/2}\left( \frac k{k_1}\right) ^{-n_1}\left[ 1-\frac{ 
(k\eta )^2}{2(2-2n_1)}\right], 
\label{outsol1}
\end{eqnarray}
where we have used the relations $\Gamma (n)\Gamma (1-n)=\pi /\sin (n\pi )$ 
and $\Gamma(-n)\Gamma (1+n)=\pi \left| 1-i\cot (\pi n)\right|$  and
defined $k_1\equiv 1/\eta _1$. Note that we
have kept, for later use, also the next to leading terms. The corrections are
of order  $(k\eta)^2$, which is a very small quantity, showing that the
approximation we use is quite good. 

The perturbation equation for $\chi^\pi$  is the following 
\begin{equation}
{\chi^\pi_k}''+\left( k^2-\frac{(S^{-1})^{\prime \prime }}{ 
S^{-1}}\right) \chi^\pi_k =0,
\label{chipert}
\end{equation}
and it's solutions are Bessel functions, as for $\chi_k$.   We normalize the
solution at early times to vacuum fluctuations, $\frac{\sqrt{k}}{\sqrt{2}}
e^{-ik\eta }$ (note the $\sqrt{k}$ instead of $1/\sqrt{k}$ in the
 normalization of the momentum). 
Substituting the explicit expressions for
$S(\eta)=a^m(\eta)e^{l\phi(\eta)/2}$, 
where $a(\eta)$ and $\phi(\eta)$ are given in
eq.(\ref{ddphi}), eq.(\ref{chipert}) takes a form similar to
eq.(\ref{earlyperteq1}), 
except that instead of  $n_s$ appears $n_s^\pi $ and instead of $n_1$ appears
$n_1^\pi $, where $n_s^\pi =n_s-1=\left( \sqrt{3}(l+m)-1-m\right) /2 $ and
$n_1^\pi =n_1-1=-\half-m$.
The solution  of eq.(\ref{chipert}) for $\eta <\eta _s$ is the following
\begin{equation}
\chi^\pi_k =\frac{\sqrt{\pi }}2\sqrt{\eta }J_{n_s^\pi }(k\eta )+
\frac{ \sqrt{\pi }}2\sqrt{\eta }Y_{n_s^\pi }(k\eta ),
\end{equation}
and for $\eta >\eta _1$ 
\begin{equation}
\chi^\pi_k =A\sqrt{\eta }J_{n_1^\pi }(k\eta )+B\sqrt{\eta } 
Y_{n_1^\pi}(k\eta ).  
\label{platesol}
\end{equation}
For the special values $n_s^\pi=\pm \half$ and $n_1^\pi=\pm \half$ the
solution reduces to a sum of exponentials. It is nevertheless possible to use
the general formulae for these special values as well.

For $\eta <\eta _s$, for perturbations outside the horizon, 
$|k\eta| \laq 1$,  $\chi^\pi$ can be approximated by
\begin{equation}
\chi^\pi_k =\frac{\Gamma \left( -n_s^\pi \right)}{2\sqrt{\pi}}
k\sqrt{\eta }\left( \frac{ k\eta }2\right)^{n_s^\pi } 
\left[ 1-\frac{ (k\eta )^2}{2(2+2n_s^\pi )}\right]  -
 i\frac{\Gamma \left( n_s^\pi \right)}{2\sqrt{\pi}}
k \sqrt{\eta }\left( \frac{k\eta } 
2\right)^{-n_s^\pi } \left[ 1-\frac{(k\eta )^2 }{2(2-2n_s^\pi )}\right], 
\end{equation}
while  for $\eta >\eta _1$, $k\eta \laq 1$, $\chi^\pi$ can be approximated by
\begin{eqnarray}
\chi^\pi_k = A_\pi \frac{\Gamma \left( -n_1^\pi \right)}{\pi}
\sqrt{\eta }\left( \frac{k\eta }2\right)^{n_1^\pi } 
\left[ 1-\frac{(k\eta )^2}{ 2(2+2n_1^\pi )}\right]  -
i B_\pi \frac{\Gamma \left( n_1^\pi \right)}{\pi}\sqrt{\eta }
\left( \frac{k\eta }2\right)^{-n_1^\pi } 
\left[ 1-\frac{(k\eta )^2}{ 2(2-2n_1^\pi )}\right]. 
\end{eqnarray}
Since  for $\eta <\eta _s$,  $\Pi=S\chi ^\pi=a_s^me^{l\phi _s/2}
\left( \eta /\eta _s\right)^{1/2-n_s}\chi^\pi$, 
\begin{eqnarray}
\Pi^{in}_k &=&\frac{2^{-1-n_s^\pi}}{\sqrt{\pi}}
\Gamma \left( -n_s^\pi \right) a_s^m e^{l\phi _s/2}
\left( k_s\right) ^{1/2}\left( \frac {k}{k_s}\right)^{n_s}
\left[ 1-\frac{(k\eta )^2}{2(2+2n_s^\pi )}\right] 
\nonumber \\
&-& i \frac{2^{n_s^\pi-1}} {\sqrt{\pi} }
\Gamma \left( n_s^\pi\right) a_s^m e^{l\phi _s/2} k 
\left( k_s\right) ^{-1/2}\left( k_s k\right)^{-n_s+1}
\eta ^{-2n_s^\pi }\left[ 1-\frac{(k\eta )^2}{2(2-2n_s^\pi )}\right], 
\label{psolin}
\end{eqnarray}
and since  for $\eta >\eta _1$, $\Pi =S\chi^\pi=
a_1^me^{l\phi _1/2}\left( \eta /\eta _1\right) ^{1/2-n_1}\chi^\pi$, 
\begin{eqnarray}
\Pi^{out}_k &=&\frac{2^{-n_1^\pi }}{\pi} A_\pi \Gamma \left( -n_1^\pi \right)
a_1^me^{l\phi _1/2}k^{-1}k_1^{1/2}\left( \frac k{k_1}\right) ^{n_1}\left[ 1- 
\frac{(k\eta )^2}{2(2+2n_1^\pi )}\right] \nonumber \\
&-&i\frac{2^{n_1^\pi }}\pi B_\pi \Gamma \left( n_1^\pi \right)
a_1^me^{l\phi _1/2}k_1^{-1/2}\left( k_1k\right) ^{-n_1+1}\eta ^{-2n_1^\pi
}\left[ 1-\frac{(k\eta )^2}{2(2-2n_1^\pi )}\right]. 
\label{psolout}
\end{eqnarray}

\subsubsection{Bridging the gap: The constant mode of $\psi$}

We will compute the contribution to the {\em out} perturbation from the
constant mode of $\psi^{in}$. 
From the second term of eq.(\ref{insol}) we obtain the the constant mode of
$\psi^{in}$
\begin{equation}
\psi^{in}_k =\frac{1}{\sqrt{\pi k_s}}2^{n_s-1}\Gamma \left( n_s\right)
a_s^{-m}e^{-l\phi _s/2}\left( \frac k{k_s}\right) ^{-n_s}_{.}  
\end{equation}
For $\eta >\eta _1$, $\psi^{out}$ is given in eq.(\ref{outsol1}).
We assume that the constant mode of $\psi$ is indeed constant during the
string  phase (and, in general, outside the horizon), as explained in the
previous section, and use this to connect the two solutions by matching them
and their first derivatives at $\eta=\eta_1$. This matching procedure is
expected to be quite accurate because we match the solutions at a point in time
at which both solutions are well outside the
horizon. We present the matching equations
once explicitly, to emphasize the importance of keeping the  next to leading
terms in eqs.(\ref{insol}) and (\ref{outsol1}).
\begin{eqnarray}
&&\frac{1}{2\sqrt{\pi k_s}}2^{n_s}\Gamma \left( n_s\right)
a_s^{-m}e^{-l\phi _s/2}\left( \frac k{k_s}\right) ^{-n_s}=
\nonumber \\  &&
B\frac {2^{n_1}}{\pi \sqrt{k_1}}\Gamma \left( n_1\right) a_1^{-m}
e^{-l\phi_1/2}\left( \frac k{k_1}\right) ^{-n_1} + 
A\frac {2^{-n_1}}{2\pi \sqrt{k_1}}\Gamma \left( -n_1\right)
a_1^{-m}e^{-l\phi _1/2}\left( k_1k\right) ^{n_1}\eta _1^{2n_1}, \nonumber \\
&& 0=-\ B \frac 2{\eta _1}\frac{(k\eta _1)^2}{2(2-2n_1)}
\frac {2^{n_1}}{\pi \sqrt{k_1} }
\Gamma \left( n_1\right) a_1^{-m}e^{-l\phi _1/2}\left( \frac
k{k_1}\right) ^{-n_1}+ \nonumber \\
&& A\frac {2^{-n_1}}{2\pi \sqrt{k_1}}\Gamma \left( -n_1\right)
a_1^{-m}e^{-l\phi _1/2}\left( k_1k\right) ^{n_1}\eta _1^{2n_1}
  \frac{2n_1}{\eta _1}_{.} 
\label{match1}
\end{eqnarray}
The solution of eq.(\ref{match1}) is 
\begin{eqnarray}
A&=&\frac{2^{n_s+n_1-2}\sqrt{\pi }}{n_1(1-n_1)}
\frac{\Gamma \left( n_s\right) }{ \Gamma \left( -n_1\right) }
z_S^{-3/2+m+n_1}\left( \frac{g_S}{g_1} 
\right) ^{-l}\left( \frac k{k_s}\right) ^{2-n_1-n_s}
\label{aeq} \\
B&=&  2^{n_s-n_1-1} \sqrt{\pi }
\frac{\Gamma \left( n_s\right) }{\Gamma\left( n_1\right) }
z_S^{1/2+m-n_1}\left( \frac{g_S}{g_1}\right) ^{-l}\left(
\frac k{k_s}\right) ^{n_1-n_s}_{.}
\label{beq} 
\end{eqnarray}
To obtain eqs.(\ref{aeq}),(\ref{beq}) we have used the definitions 
$z_S=a_1/a_S$, $g_1=e^{\phi_1/2}$, $g_S=e^{\phi_S/2}$.
Note that had we not kept the next to leading terms in eq.(\ref{insol}) 
we would have obtained $A=0$.

To summarize, plugging in the resulting coefficients from eqs.(\ref{aeq}) and
(\ref{beq}) into eq.(\ref{outsol1}) we obtain 
\begin{equation}
\psi_k^{out}= 
\frac {1}{\sqrt{\pi k_s}}
2^{n_s-1}\Gamma \left( n_s\right) 
a_s^{-m}g_S^{-l}z_S^{2m}\left( \frac k{k_s}\right)^{-n_s} +
  \frac 1{\sqrt{\pi k_s}}
\frac{2^{n_s-2}\Gamma \left( n_s\right) }{ n_1(1-n_1)}
a_s^{-m}g_S^{-l} z_S^{n_1-1}\left( \frac{k}{k_s}\right)^{2-n_s}
(k_s\eta )^{2n_1}_{.}
\label{outsol2}
\end{equation}

The $out$ solution (\ref{outsol2}) may be put, using the 
the relation $k_S z_S=k_1$, in the 
form $\psi_k^{out}=C_1\left[1+ C_2
\left(\frac{k}{k_1}\right)^2 (k_1 \eta)^{2n_1}\right]$, where $C_2$ is a
numerical coefficient of order 1. At $\eta=\eta_1$, the ratio of
the two terms is approximately $(k\eta_1)^2\ll 1$, because $k_1\eta_1=1$.
However, at later times, using eq.(\ref{latesol}), the general solution  
is given by 
$\psi^{out}_k \simeq  A\ S^{-1}(\eta) \sqrt{\eta }J_{n_1}(k\eta ) +
B\ S^{-1}(\eta)\sqrt{\eta }Y_{n_1^\pi}(k\eta )$. 
A good time to look at is near $\eta_{re}$. At that time the magnitude of both
Bessel functions is generically the same, and to decide on the leading
contribution we look at the ratio of the coefficients $A/B\simeq
\left(\frac{k}{k_1}\right)^{2-2n_1}$.
Therefore the leading contribution at much later times depends on $n_1$. For
$n_1>1$, $A\gg B$ and the leading contribution will be from the time dependent
term in eq.(\ref{outsol2}), and for $n_1<1$, $B\gg A$ and the leading
contribution will be from the  constant term in eq.(\ref{outsol2}).  In
conclusion, for later times,
\begin{eqnarray}
\psi^{out}_k&\simeq& A\ S^{-1}(\eta) \sqrt{\eta }J_{n_1}(k\eta ) 
\hspace{.5in}  n_1>1 
\label{bessolA} \\
\psi^{out}_k&\simeq& B\ S^{-1}(\eta)\sqrt{\eta }Y_{n_1}(k\eta )
\hspace{.5in}  n_1<1,
\label{bessolB}
\end{eqnarray}
where $A$ and $B$ are given by eq.(\ref{aeq}) and eq.(\ref{beq}).

\subsubsection{Bridging the gap: The time dependent mode of $\psi$ }

We will compute the contribution to the {\em out} perturbation from the
time dependent mode of $\psi^{in}$.  The time dependent mode of $\psi$ can
be extracted from the first term in eq.(\ref{insol}),
\begin{equation}
\psi_k^{in} =\frac{1}{2\sqrt{\pi k_s}}2^{-n_s}\Gamma \left( -n_s\right)
a_s^{-m}e^{-l\phi _s/2}\left( k_sk\right) ^{n_s}\eta ^{2n_s}_{.}
\end{equation}
Since for $\eta <\eta _s,$ $S^2(\eta)=
a_s^{2m}e^{l\phi _s}\left( \eta /\eta_s\right) ^{1-2n_s}$, we obtain 
\begin{equation}
{{\psi}_k^{in}}' S^2=\frac{n_s\sqrt{k_s}}{\sqrt{\pi }}2^{-n_s}\Gamma \left(
-n_s\right) a_s^me^{l\phi _s/2}\left( \frac k{k_s}\right) ^{n_s}=const.
\end{equation}
We can see explicitly that the time dependent mode of $\psi $ corresponds to
the constant mode of the canonically conjugate momentum $\Pi$. 

We compute the contribution from the time dependent mode of $\psi$ using the
constant mode of $\Pi$, assuming that it is constant throughout the string
phase (and, in general, outside the horizon), as explained in the previous
section, and use this to connect the two solutions by matching them and their
first derivatives at $\eta=\eta_1$. This matching procedure is expected to be
quite accurate because we match the solutions at a point in time at which both
solutions are well outside the horizon.

The solutions are given in eq.(\ref{psolin}) and (\ref{psolout})
and  their matching yields
\begin{eqnarray}
A_\pi &=& \sqrt{\pi } 2^{n_1-n_s-1}
\frac{\Gamma \left(1 -n_s \right)}{\Gamma \left( 1-n_1 \right) }
z_S^{-1/2-m+n_1}\left( \frac{g_S}{g_1}\right) ^l
k \left( \frac {k}{k_s}\right)^{-n_1+n_s}
\label{api} \\
B_\pi &=&\sqrt{\pi }
\frac {2^{3n_1-n_s-2}}{n_1(1-2n_1)}
\frac{\Gamma \left(1-n_s \right) }{\Gamma \left( n_1-1\right) } 
z_S^{-1/2-m-n_1}\left( \frac{g_S}{g_1}\right) ^l
k \left( \frac k{k_s}\right)^{n_1+n_s}, 
\label{bpi}
\end{eqnarray}
where we have used the definitions of $n_s^\pi$ and $n_1^\pi$.
Substituting into eq.(\ref{psolout}) we obtain
\begin{equation}
\Pi^{out}_k =\frac{2^{1-n_1 }}{\pi} A_\pi 
\Gamma \left( 1-n_1 \right)
a_1^me^{l\phi _1/2}k^{-1}k_1^{1/2}\left( \frac k{k_1}\right) ^{n_1}
 -
 i \frac{2^{n_1-1}}\pi B_\pi \Gamma \left( n_1-1\right)
a_1^me^{l\phi _1/2}k_1^{-1/2}\left( k_1k\right) ^{-n_1+1}\eta ^{2-2n_1}
\label{outsol3}
\end{equation}
with $A_\pi$ and $B_\pi$ given in eq.(\ref{api}) and (\ref{bpi}).

As for the case of the constant mode of $\psi$, we may compare the relative
strength of the two terms. 
At $\eta=\eta_1$, the relative strength of
the two terms is approximately $(k\eta_1)^2\ll 1$, because $k_1\eta_1=1$.
However, at later times, using eq.(\ref{platesol}), the general solution  
is given by 
$\Pi^{out}_k \simeq  A_\pi\ S(\eta) \sqrt{\eta }J_{n_1^\pi} (k\eta ) +
B_\pi\ S(\eta)\sqrt{\eta }Y_{n_1^\pi} (k\eta )$. 
A good time to look at is near $\eta_{re}$. At that time the magnitude of both
Bessel functions is generically the same and to decide on the leading
contribution we look at the ratio of the coefficients $A_\pi/B_\pi\simeq
\left(\frac{k}{k_1}\right)^{-2n_1}$.
Therefore the leading contribution at much later times depends on $n_1$. For
$n_1>0$, $A_\pi\gg B_\pi$ and the leading contribution will be from the time
dependent term in eq.(\ref{outsol3}), and for $n_1<0$, $B_\pi\gg A_\pi$ and the
leading contribution will be from the  constant term in eq.(\ref{outsol3}). 
In conclusion, for later times,
\begin{eqnarray}
\Pi_k^{out}&\simeq&S(\eta)\ A_\pi \sqrt{\eta }J_{n_1^\pi }(k\eta ) 
\hbox{ for } n_1>0\nonumber \\
\Pi_k^{out}&\simeq& S(\eta)\ B_\pi \sqrt{\eta }Y_{n_1^\pi }(k\eta ) 
\hbox{ for } n_1<0
\label{bessol2}
\end{eqnarray}
where $A_\pi$ and $B_\pi$ are given in eqs.(\ref{api}), (\ref{bpi}) .

\subsection{Late time solutions and spectrum: The massless approximation}

The perturbation equation in the range  $\eta >\eta _{re}$ is  $
\chi''_k+\left(k^2+M^2a^2-\frac{S''}{S}\right) \chi_k =0$.  Because we cannot
solve this equation in a closed form for the general case, we use an
approximate form of the equation to obtain approximate solutions. We
distinguish between two cases,  $k>Ma_{re}$ - the ``massless approximation'';
$k<Ma_{re}$ - the ``massive approximation'', where $a_{re}=a(\eta_{re})$. 

In the massless approximation the perturbation
equation is  $\chi''_k+\left(k^2-\frac{S''}{S}\right) \chi_k =0$, which is the
same equation we solved already, so there's no need for any additional steps.
All we need to do is to take the late time solutions and interpret them. In a
standard way, we compute from the solution  the number of produced particles at
a given wave number $k$, $N_k$,  and the ratio of the energy density per octave
of the produced particles to the critical energy density
$\frac{d\Omega}{d\ln{\omega}}$.  For massless particles  
$d\Omega/(d\ln\omega)=w^4 N_\omega /(M_p H_0)^2$,  
where $H_0$ is today's Hubble parameter,
$M_p$ is the Planck mass and $\omega=k/a(\eta)$ \cite{comm}. 

We can further simplify  the denominator $(M_p H_0)^2$. For that we switch
momentarily to using cosmic time. We may reexpress the relation
$H_0=H_{eq}\frac{t_{eq}}{t_0}$,  as 
$H_0=H_1\frac{t_1}{t_{eq}}\frac{t_{eq}}{t_0}$, where $H_1$ is the Hubble
parameter at $t_1$, the onset of RD.
In MD $a(t)=a_{eq}\left( \frac {t}{t_{eq}}\right)^{2/3}$
and therefore $\frac{t_{eq}}{t_0}=\left( \frac {a_0}{a_{eq}}\right) ^{-3/2}$.
In RD
$a_{eq}=a_1\left( \frac{t_{eq}}{t_1}\right)^{1/2}$, then 
$\frac{t_1}{t_{eq}}=\left( \frac{a_{eq}}{a_1}\right)^{-2}$.
Combining these relations, we obtain 
$
H_0=H_1\left( \frac{a_{eq}}{a_1}\right)^{-2}
\left(\frac{a_0}{a_{eq}}\right)^{-3/2}$.
Since we assumed that $H_1=M_s$ then 
$H_0=M_s\left( \frac{a_1}{a_0}\right)^2
\left( \frac {a_0}{a_{eq}}\right)^{1/2}_{.}$
Now,  $M_p=M_s g_1^{-1}$, $M_s=\omega_1 a_0/a_1$ and
$\omega_1=\omega_s z_S$, and therefore 
$M_s \frac{a_1}{a_0}=\omega_s z_S$.  Combining these relations we arrive at our
final destination
\begin{eqnarray}
M_p H_0&=&M_s^2\left( \frac{a_1}{a_0}\right)^2
z_{eq}^{1/2} g_1^{-1}\nonumber \\ 
&=& \omega_s^2 z_S^2 z_{eq}^{1/2} g_1^{-1},
\label{denom}
\end{eqnarray}
where $z_{eq}=a_0/a_{eq}$.

Substituting eq.(\ref{denom}) into the expression for the energy density we
obtain the following expression
\begin{equation}
d\Omega /(d\ln\omega)=z_{eq}^{-1}z_S^{-4}g_1^2
\left(\frac{\omega}{\omega_s}\right)^4 N_\omega. 
\label{odef}
\end{equation}
The factor  $z_{eq}^{-1}$, the redshift from the time of matter-radiation
equality and today, appears because the energy density
of massless particles redshits faster than that of matter.

We may also discuss spectra of massive particles in the massless
approximation. Although this may sound a little strange, such an approximation
does make sense, because it is made at $\eta_{re}$, when the effects of the
mass can be negligible. The only difference between massive particles and
massless particles in the massless approximation is in their energy density. 
The solutions  near $\eta_{re}$ are approximately the same, and
therefore the number of produced particles is the same as in the massless
case. However, at much later times, the contribution from the mass may
dominate the energy density. We can then compute  the ratio of the energy
density per octave of the produced particles to the critical energy density
$\frac{d\Omega}{d\ln{\omega}}$.  For massive particles   $d\Omega
/(d\ln\omega)=\sqrt{M^2+\omega^2} w^3 N_\omega /(M_p H_0)^2$,  where $H_0$ is
today's Hubble parameter, $M_p$ is the Planck mass and $\omega=k/a(\eta)$.
Using eq.(\ref{denom}) we may show that 
\begin{equation} 
d\Omega/(d\ln\omega)=z_{eq}^{-1}z_S^{-4}g_1^2
\frac{\sqrt{M^2+\omega^2}}{\omega_s}
\left(\frac{\omega}{\omega_s}\right)^3 N_\omega. 
\label{modef} 
\end{equation}

An important difference compared with the massless spectrum is that one power
of $\omega/\omega_s$ is replaced by $\frac{\sqrt{M^2+\omega^2}}{\omega_s}$
(for $\omega<M$). Because the energy density of massive particles redshits
slower than that of radiation, two enhancement factors will appear compared
with  the case of massless particles. One enhancement factor accounts for the
period between reentry and matter-radiation equality, in which the relative
energy density of massive particles grows, while that of massless particles
remains constant. Another enhancement factor appears because the relative
energy density of massive particles stays constant during the matter dominated
era, while that of massless particles decreases. It is harder to see this
factor in the equations  because $z_{eq}$  also appears in the  expressions
for massive particles. However, it's appearance is artificial. The combination
$M/(z_{eq} \omega_s)=M/\omega_s(\eta_{eq})$ actually does not  depend on the
redshift from matter-radiation equality epoch until today.

\subsubsection{Massless spectrum: contribution of the constant mode of $\psi$}

The approximate late time solutions of the perturbation equation are given in
eqs.(\ref{bessolA}), (\ref{bessolB}). 
Using the asymptotic expansion of Bessel functions, 
the solutions (either during RD or MD) approach, for $k\eta> 1$
\begin{equation}
\psi_k =S^{-1}\frac 1{\sqrt{2k}}\left( \alpha _ke^{-ik\eta
}+\beta _ke^{ik\eta }\right) 
\end{equation}
where 
\begin{equation}
\left| \alpha _k\right| =\left| \beta _k\right| =2^{n_s-n_1}
\left| \frac{\Gamma \left(n_s\right) }{\Gamma \left( n_1\right) } \right|
z_S^{1/2+m-n_1}\left( \frac{g_S }{g_1}\right)^{-l}
\left( \frac k{k_s}\right) ^{n_1-n_s}\hbox{ for }n_1<1
\end{equation}
and
\begin{equation}
\left| \alpha _k\right| =\left| \beta _k\right| =
\frac {2^{n_s+n_1-1}}{n_1(1-n_1)}
\left| \frac{ \Gamma \left( n_s\right) }{\Gamma \left( -n_1\right) } \right|
z_S^{-3/2+m+n_1}\left( \frac{g_S}{g_1}\right) ^{-l}\left( \frac
k{k_s}\right) ^{2-n_1-n_s}\hbox{ for }n_1>1
\end{equation}

The number of massless particles $N_k$, can be
expressed in the standard form $N_k=\left|\beta_k\right|^2$.  
Using eq.(\ref{odef}), $n_1=\half-m$, and using  
$d\Omega_\psi$ to denote the contribution from the constant mode of $\psi$
we arrive at the final expressions 
\begin{eqnarray}
\left[ \frac{d\Omega _\psi }{d\ln \omega}\right] _{n_1<1}&=&
2^{2n_s+2m-1}
\frac{\Gamma^2\left( n_s\right) }{\Gamma ^2\left(\half-m\right) } 
z_{eq}^{-1} g_1^2 z_S^{4m-4}\left(\frac{g_S}{g_1}\right)^{-2l}
\left( \frac{\omega}{\omega_s}\right)^{5-2m-2n_s}
\label{psinl1} \\
\left[ \frac{d\Omega _\psi }{d\ln \omega}\right] _{n_1>1}&=&
\frac {2^{2n_s-2m-1}}{(\quarter-m^2)^2} 
\frac{\Gamma ^2\left( n_s\right) }{\Gamma^2\left(m-\half\right) }
z_{eq}^{-1} g_1^2 z_S^{-6}\left( \frac{g_S}{g_1}\right)^{-2l}
\left( \frac {\omega}{\omega_s}\right)^{7+2m-2n_s}_{.}
\label{psing1}
\end{eqnarray}

\subsubsection{Massless spectrum: contribution of the time dependent mode of
$\psi$}

As usual, we use the constant mode of $\Pi$ to evaluate the contribution of the
time dependent mode of $\psi$. 
The approximate late time solutions of the perturbation equation are given in
eq.(\ref{bessol2}). Using the asymptotic expansion of Bessel functions, 
the solutions (either during RD or MD) approach, for $k\eta> 1$
\begin{equation}
\Pi_k =S \frac {\sqrt{k}}{\sqrt{2}}\left( \alpha _ke^{-ik\eta}+
\beta _ke^{ik\eta }\right) 
\end{equation}
where 
\begin{equation}
\left| \alpha _k\right| =\left| \beta _k\right| =2^{n_1-n_s}
\left| \frac{\Gamma\left(-n_s^\pi\right) }{\Gamma\left(-n_1^\pi\right) }\right| 
z_S^{-1/2-m+n_1}\left( \frac{g_S}{g_1}\right) ^l
\left( \frac k{k_s}\right)^{-n_1+n_s}\hbox{ for }n_1>0
\end{equation}
and
\begin{equation}
\left| \alpha _k\right| =\left| \beta _k\right| =
\frac {2^{3n_1-n_s-1}}{|n_1(1-2n_1)|}
\left| \frac{ \Gamma \left( -n_s^\pi \right) }
{\Gamma \left( n_1^\pi\right) } \right|
z_S^{-1/2-m-n_1}\left( \frac{g_S}{g_1}\right)^l
\left( \frac{k}{k_s}\right) ^{n_1+n_s}\hbox{ for }n_1<0.
\end{equation}
The number of massless particles $N_k$ of a given wavenumber  $k$, can be
expressed as $N_k=\left|\beta_k\right|^2$. Note that we are calculating the
number of particles directly from the solution for $\Pi_k$. We could have also
used the relation $\Pi=S^2\psi'$, use the standard definition of $N_k$ from
the coefficients in the solution for $\psi$ and obtain the same result. 

Using eq.(\ref{odef}), $n_1=\half-m$, and using  
$d\Omega_\pi$ to denote the contribution from the time dependent mode of $\psi$
(the constant  mode of $\Pi$)
we arrive at the final expressions 
\begin{eqnarray}
\left[ \frac{d\Omega _\pi }{d\ln \omega}\right] _{n_1>0}&=&2^{1-2m-2n_s}
\frac{\Gamma^2\left(1 -n_s \right) }{\Gamma ^2\left( m+\half \right) } 
z_{eq}^{-1} g_1^2 z_S^{-4m-4} \left( \frac{g_S}{g_1}\right) ^{2l}
\left( \frac{\omega}{\omega_s}\right)^{3+2m+2n_s}
\label{ping0}\\
\left[ \frac{d\Omega _\pi }{d\ln \omega}\right] _{n_1<0}&=&
\frac{2^{-6m -2n_s-5}}{m^2(m-\half)^2}
\frac{\Gamma ^2\left( 1-n_s \right)}{\Gamma ^2\left( -\half -m \right) }
z_{eq}^{-1} g_1^2 z_S^{-6} \left( \frac{g_S}{g_1}\right) ^{2l}
\left( \frac {\omega}{\omega_s}\right)^{5-2m+2n_s}_{.}
\label{pinl0}
\end{eqnarray}

\subsubsection{Massive spectrum: contribution of the constant mode of
$\psi$}

The final result for the contribution of the constant mode of $\psi$ to the
spectrum of massive particles (recall that we are using the massless approximation, defined at the
beginning of this section) maybe read off from the final result for
massless particles, eqs.(\ref{psinl1}) and (\ref{psing1}), by making the
substitution of one power of $\omega/\omega_s$ by one power of 
$\frac{\sqrt{M^2+\omega^2}}{\omega_s}$,
\begin{eqnarray}
\left[ \frac{d\Omega _\psi }{d\ln \omega}\right] _{n_1<1}&=&
2^{2n_s+2m-1}
\frac{\Gamma^2\left( n_s\right) }{\Gamma ^2\left(\half-m\right) } 
z_{eq}^{-1} g_1^2 z_S^{4m-4}\left(\frac{g_S}{g_1}\right)^{-2l}
\left( \frac{\omega}{\omega_s}\right)^{4-2m-2n_s}
\frac{\sqrt{M^2+\omega^2}}{\omega_s}
\label{psinl1mass} \\
\left[ \frac{d\Omega _\psi }{d\ln \omega}\right] _{n_1>1}&=&
\frac {2^{2n_s-2m-1}}{(\quarter-m^2)^2} 
\frac{\Gamma ^2\left( n_s\right) }{\Gamma^2\left(m-\half\right) }
z_{eq}^{-1} g_1^2 z_S^{-6}\left( \frac{g_S}{g_1}\right)^{-2l}
\left( \frac {\omega}{\omega_s}\right)^{6+2m-2n_s}
\frac{\sqrt{M^2+\omega^2}}{\omega_s}.
\label{psing1mass}
\end{eqnarray}

\subsubsection{Massive spectrum: contribution of the time dependent mode of
$\psi$}

The final result for the contribution of the constant mode of $\psi$ to the
spectrum of massive particles (recall that we are using the massless approximation, defined at the
beginning of this section) maybe read off from the final result for
massless particles, eqs.(\ref{ping0}) and (\ref{pinl0}), by making the
substitution of one power of $\omega/\omega_s$ by one power of 
$\frac{\sqrt{M^2+\omega^2}}{\omega_s}$,
\begin{eqnarray}
\left[ \frac{d\Omega _\pi }{d\ln \omega}\right] _{n_1>0}&=&2^{1-2m-2n_s}
\frac{\Gamma^2\left(1 -n_s \right) }{\Gamma ^2\left( m+\half \right) } 
z_{eq}^{-1} g_1^2 z_S^{-4m-4} \left( \frac{g_S}{g_1}\right) ^{2l}
\left( \frac{\omega}{\omega_s}\right)^{2+2m+2n_s}
\frac{\sqrt{M^2+\omega^2}}{\omega_s}
\label{ping0mass}\\
\left[ \frac{d\Omega _\pi }{d\ln \omega}\right] _{n_1<0}&=&
\frac{2^{-6m -2n_s-5}}{m^2(m-\half)^2}
\frac{\Gamma ^2\left( 1-n_s \right)}{\Gamma ^2\left( -\half -m \right) }
z_{eq}^{-1} g_1^2 z_S^{-6} \left( \frac{g_S}{g_1}\right) ^{2l}
\left( \frac {\omega}{\omega_s}\right)^{4-2m+2n_s}
\frac{\sqrt{M^2+\omega^2}}{\omega_s}.
\label{pinl0mass}
\end{eqnarray}

\subsection{Late time solutions and spectrum: The massive approximation}

If the produced particles are massive, and at $\eta_{re}$, $M a_{re}>k$, we are
unable to solve the late time equation exactly, and therefore need to resort to
another matching procedure. We choose to perform the matching at
$\eta=\eta_{re}$, where $k^2+M^2 a_{re}^2\sim H^2$. We match a solution from
outside the horizon to a massive solution inside. This is not an ideal match,
but both are expected to be fairly accurate at $\eta_{re}$. Note that the
solutions outside the horizon were obtained using a ``massless" equation,
however outside the horizon the dominant term is $S''/S$ anyhow, and therefore
no large inaccuracies should be induced. We then need to know whether
$\eta_{re}$ occurs during RD or MD. If $M a_{eq}>H(\eta_{eq})$ then the the
reentry time of all perturbation will be at earlier times. Therefore, for $M>
10^{-27}eV$, the reentry time is during RD. Since we do not expect any
particles to have masses smaller than $10^{-27}eV$, we continue under the
assumption that $\eta_{re}<\eta_{eq}$.  We have performed the calculations for
the case $\eta_{re}>\eta_{eq}$, however, we will not present them in this
paper. 

In a standard way,
we compute from the solution  the number of produced particles at a given wave
number $k$, $N_k$.  We then compute 
the ratio of the energy density per octave of the
produced particles to the critical energy density
$\frac{d\Omega}{d\ln{\omega}}$.  As already shown, for massive particles  
$d\Omega /(d\ln\omega)$ is given by eq.(\ref{modef}).

The equation of motion in the massive approximation is 
\begin{equation}
\chi_k^{\prime \prime }+M^2a^2\chi_k =0
\end{equation}
and it's solutions in the WKB approximation are given by
\begin{equation}
\chi_k^{\pm} =
\frac 1{\sqrt{2Ma(\eta )}}\exp \left( \pm i\int^\eta Ma(\eta ^{\prime
})d\eta ^{\prime }\right) ,
\end{equation}
where for the RD epoch ($\eta <\eta _{eq})$ 
\begin{equation}
\chi_k^{\pm} =
\frac 1{\sqrt{2Ma_1k_1\eta }}\exp \left( \pm i\frac 12Ma_1k_1\eta^2\right),
\end{equation}
Since $\psi =S^{-1}\chi$ the general solution for the massive perturbation for
RD is the following,
\begin{equation}
\psi^{massive}_k=
\frac {1}{\sqrt{2Ma_1k_1\eta }}S^{-1}(\eta )\left[ \alpha _k\exp
\left( -i\frac 12Ma_1k_1\eta ^2\right) +\beta _k\exp \left( i\frac
12Ma_1k_1\eta ^2\right) \right].
\label{massRD} 
\end{equation}
We will also need the general solution for the conjugate momentum, which can
either be derived form the equation for $\chi^\pi$ or from the relation
$\Pi=S^2\psi'$, both yielding the approximate solution
\begin{equation}
\Pi^{massive}_k=
{\sqrt{2Ma_1k_1\eta }}S(\eta )\left[ \alpha^\pi_k\exp
\left( -i\frac 12Ma_1k_1\eta ^2\right) +\beta^\pi_k\exp \left( i\frac
12Ma_1k_1\eta ^2\right) \right].
\label{pmassRD} 
\end{equation}
Instead of using only $\psi$ at late times, we will again use $\Pi$,
making sure that we do not double count the number of produced particles.

\subsubsection{Massive spectrum: The contribution of the constant mode of
$\psi$}

As in the massless case, we assume that $H_1^2>k^2+M^2a_1^2,$ for all $k,$
therefore solutions for $\eta>\eta_1$ are given in eqs.(\ref{bessolA}), 
(\ref{bessolB}). We reproduce here
their approximate forms,
\begin{eqnarray}
\psi_k^{out}&=&S^{-1}(\eta)\frac{\sqrt{\pi }2^{n_s+n_1-2}}{n_1(1-n_1)}
\frac{\Gamma\left( n_s\right) }{\Gamma \left( -n_1\right) }
z_S^{-1}\left( \frac{g_S}{g_1}\right)^{-l}
\left( \frac k{k_s}\right)^{2-n_1-n_s}\!\!\sqrt{\eta } J_{n_1}(k\eta )
\hbox{ for } {n_1>1}
\\
\psi_k^{out}&=&S^{-1}(\eta) \sqrt{\pi } 2^{n_s-n_1-1} 
\frac{\Gamma \left( n_s\right) }{\Gamma \left( n_1\right) }
z_S^{2m}\left( \frac{g_S}{g_1}\right)^{-l}
\left( \frac k{k_s}\right) ^{n_1-n_s}\!\!\sqrt{\eta }J_{-n_1}(k\eta )
\hbox{ for } {n_1<1}
\end{eqnarray}
In the case of the massive approximation these solutions are correct only for 
$ H^2>k^2+M^2a^2$. Therefore for $k\eta < 1$ (and $M<k/a)$ the leading 
terms are
\begin{equation}
\psi _{n_1>1}=\frac 1{\sqrt{\pi k_s}}
\frac{2^{n_s-2}\Gamma \left( n_s\right) }{ n_1(1-n_1)}
a_s^{-m}z_S^{2(n_1-1)}g_S^{-l}\left( \frac{k}{k_s}\right)^{2-n_s}
(k_s\eta )^{2n_1}
\label{tdnge1}
\end{equation}
for the time dependent mode, and 
\begin{equation}
\psi _{n_1<1}=\frac 1{\sqrt{\pi k_s}}
2^{n_s-1}\Gamma \left( n_s\right) 
a_s^{-m}g_S^{-l}\left( \frac k{k_s}\right)^{-n_s}
\label{tdnle1}
\end{equation}
for the constant mode.

{\em (a) The case $n_1>1$.}\\
In the case $n_1>1$, the approximate solution for $\eta \laq \eta _{re}$ 
is given in eq.(\ref{tdnge1}), 
and for $\eta \gaq \eta _{re}$ we obtained the approximate solution 
in eq.(\ref{massRD}). 
{}We connect the solutions at $\eta=\eta_{re}$ by matching them and 
their first derivatives,  
$\psi_k^{out}(\eta _{re})=\psi_k^{massive}(\eta _{re})$ and
${\psi_k^{out}}(\eta _{re})'={\psi_k^{massive}}(\eta _{re})'$.
Using the relations $H(\eta _{re})=Ma(\eta _{re})$ and 
$\eta _{re}^{-1}\equiv k_{re}=\left(Ma_1k_1\right)^{1/2}$, 
the results of the matching give
\begin{equation}
\left| \beta _k\right| =\frac{\left| 3m+i\right|}{\sqrt{ \pi }}
\frac{2^{n_s-3}
\left|\Gamma \left( n_s\right)\right| }{ |n_1(1-n_1)|}z_S^{-1}\left( 
\frac{g_S}{g_1}\right) ^{-l}\left( \frac k{k_s}\right)^{2-n_s}
\left( \frac{ k_{re}}{k_s}\right)^{-n_1}.
\end{equation}

Since $k_{re}/k_s=\left( Ma_1/k_1\right) ^{1/2}z_S$, and since we assume as in
\cite{peak} $k_1/a_1=M_s$, we obtain
\begin{equation}
\left| \beta _k\right|=
\frac{\left| 3m+i\right|} {\sqrt{\pi }}
\frac{2^{n_s-3} \left|\Gamma \left( n_s\right)\right| }{|n_1(1-n_1)|} 
z_S^{m-3/2}
\left( \frac{g_S}{g_1}\right) ^{-l}
\left( \frac {\omega}{\omega_s}\right) ^{2-n_s}
\left( \frac{M}{M_s}\right) ^{-n_1/2}_{.} 
\end{equation}

Using eq.(\ref{modef}), and the definition of $n_1$,
we obtain the final result
\begin{equation}
\left[ \frac{d\Omega _\psi }{d\ln \omega}\right] _{n_1>1}=
\frac{\left|3m+i\right| ^2}{\pi }
\frac{2^{2n_s-6} }{ (\quarter-m^2)^2} \Gamma ^2\left( n_s\right) 
 z_{eq}^{-1} g_1^2 z_S^{2m-7}\left( \frac{g_S}{g_1}\right) ^{-2l}
\frac{\sqrt{M^2+\omega^2}}{\omega_s}\left( \frac {\omega}{\omega_s}\right)^{7-2n_s}
\left( \frac{M}{M_s}\right)^{m-1/2}_{.} 
\label{mrdng1}
\end{equation}

{\em (b) The case $n_1<1$.}\\
In the case $n_1<1$ we obtained the approximate form of the solution
 for $\eta \laq \eta _{re}$ in eq.(\ref{tdnle1}).
and for $\eta \gaq \eta _{re}$ we obtained the approximate form of the
solution in eq.(\ref{massRD}).
We connect the solutions at $\eta=\eta_{re}$ by matching them and 
their first derivatives,  
$\psi_k^{out}(\eta _{re})=\psi_k^{massive}(\eta _{re})$ and
${\psi_k^{out}}(\eta _{re})'={\psi_k^{massive}}(\eta _{re})'$.
Using the relations $H(\eta _{re})=Ma(\eta _{re})$ and 
$\eta _{re}^{-1}\equiv k_{re}=\left(Ma_1k_1\right) ^{1/2}$, 
the results of the matching give
\begin{equation}
\left| \beta _k\right| =\frac{\left| \half+m+i\right| } {\sqrt{\pi }} 
2^{n_s-2}\left|\Gamma \left( n_s\right)\right| z_S^{2m}
\left( \frac{g_S}{g_1}\right)^{-l}\left( \frac k{k_s}\right) ^{-n_s}
\left( \frac{k_{re}}{k_s}\right)^{n_1}_{.}
\end{equation}
Since $k_{re}/k_s=\left( Ma_1/k_1\right) ^{1/2}z_S$ and since we assume as in
\cite{peak} $k_1/a_1=M_s$ we obtain 
\begin{equation}
\left| \beta _k\right|=\frac{\left| \half+m+i\right| } 
{\sqrt{\pi }} 2^{n_s-2}\left|\Gamma \left( n_s\right)\right| z_S^{m+1/2}
\left( \frac{g_S}{g_1}\right) ^{-l}
\left( \frac {\omega}{\omega_s}\right)^{-n_s}
\left( \frac{M}{M_s}\right) ^{n_1/2}_{.} 
\end{equation}
Using eq.(\ref{modef}), and the definition of $n_1$, 
we obtain the final result
\begin{equation}
\left[ \frac{d\Omega _\psi }{d\ln \omega}\right]_{n_1<1}\hspace{-.3in}=
\frac{\left|\half+m+i\right| ^2}{\pi }
2^{2n_s-4}\Gamma ^2\left( n_s\right)
z_{eq}^{-1} g_1^2 z_S^{2m-3}\left( \frac{g_S}{g_1}\right)^{-2l}
\frac{\sqrt{M^2+\omega^2}}{\omega_s}
\left( \frac {\omega}{\omega_s}\right)^{3-2n_s}
\left( \frac{M}{M_s}\right)^{1/2-m}_{.} 
\label{mrdnl1}
\end{equation}
Note that the spectrum is actually partially amplified because the
relative density of massive particles during RD is growing as the scale factor.

\subsubsection{Massive spectrum: The contribution of the time dependent
 mode of $\psi$}

As in the massless approximation, we assume that $H_1^2>k^2+M^2a_1^2,$ for
all $k$. We use our method of calculating the contribution of the
time dependent mode of $\psi$ by using it's correspondence to the constant
mode of $\Pi$.  The  solutions are given in
eq.(\ref{bessol2}). We reproduce here their approximate forms, 
\begin{eqnarray}
\Pi_k^{out} &=& S(\eta)\sqrt{\pi } 2^{n_1-n_s-1 }
\frac{\Gamma \left( -n_s^\pi\right) }{\Gamma \left( -n_1^\pi \right) }
z_S^{-2m} 
\left( \frac{g_S}{g_1}\right)^l\left( \frac {k}{k_s}\right)^{-n_1+n_s}
\!\!k\ \sqrt{ \eta }J_{n_1^\pi }(k\eta ) \hbox{ for }  n_1>0 
\label{apppia}\\
\Pi_k^{out} &=& S(\eta)\frac{\sqrt{\pi} 2^{3n_1-n_s-2}}{ n_1(1-2n_1)}
\frac{\Gamma\left( -n_s^\pi \right) }{\Gamma \left( n_1^\pi \right) } 
z_S^{-1}\left( \frac{g_S}{g_1}\right)^l
\left( \frac {k}{k_s}\right)^{n_1+n_s}
\!\! k\ \sqrt{\eta}J_{-n_1^\pi }(k\eta ) \hbox{ for } n_1<0.
\label{apppib}
\end{eqnarray}
In the case of massive particles these solutions are correct only for 
$H^2>k^2+M^2a^2$.
Therefore  for $k\eta < 1$ (and $M<k/a)$ the leading terms are 
\begin{equation}
\Pi_{k}^{out}=\sqrt{\pi k_s} 2^{1-n_s^\pi }
\Gamma \left( -n_s^\pi\right) 
a_s^mg_S^l\left( \frac k{k_s}\right) ^{n_s} \hbox { for } n_1>0,
\label{pnge0}
\end{equation}
and
\begin{equation}
\Pi_{k}^{out}= \sqrt{\pi k_s} 
\frac{2^{4n_1-n_s-3}}{ n_1 (1-2n_1)} \Gamma \left( -n_s^\pi\right) 
a_s^m g_S^lz_S^{-1+2m}\left( \frac
k{k_s}\right) ^{1+n_s-2m}(k\eta )^{2m+1} \hbox { for } n_1<0.
\label{pnle0}
\end{equation}

{\em (a)  The case $n_1>0$.}\\
In the case $n_1>0$, the approximate solution for $\eta \laq\eta _{re}$ is
given in eq.(\ref{pnge0})  
and for $\eta \gaq \eta _{re}$ it is given in eq.(\ref{pmassRD}).
We connect the solutions at $\eta=\eta_{re}$ by matching them and 
their first derivatives,  
$\Pi_k^{out}(\eta _{re})=\Pi_k^{massive}(\eta _{re})$ and
${\Pi_k^{out}}(\eta _{re})'={\Pi_k^{massive}}(\eta _{re})'$,
and obtain
\begin{equation}
\left|\beta _k\right|=\left|\half+m+i\right|2^{-2-n_s^\pi}\sqrt{\pi} 
\left|\Gamma \left( -n_s^\pi \right)\right| z_S^{-2m}\left( 
\frac{g_S}{g_1}\right)^l\left( \frac k{k_s}\right) ^{n_s}\left( \frac{k_{re} 
}{k_s}\right) ^{-n_1}_{.}
\end{equation}
Using $k_{re}/k_s=\left( Ma_1/k_1\right) ^{1/2}z_S$ and since we assume as in
\cite{peak} $k_1/a_1=M_s$, 
\begin{equation}
\left|\beta_k\right|=\left|\half+m+i\right| 
2^{-2-n_s^\pi}\sqrt{\pi}
\left|\Gamma \left( -n_s^\pi \right)\right| z_S^{-m-1/2}
\left( \frac{g_S}{g_1}\right) ^l\left( \frac {\omega}{\omega_s}\right) ^{n_s}
\left( \frac{M}{M_s}\right)^{-n_1/2}_{.} 
\end{equation}
Using eq.(\ref{modef}), and the definitions of $n_s^\pi$ and
$n_1$, we get the final result
\begin{equation}
\left[ \frac{d\Omega _\pi }{d\ln \omega}\right] _{n_1>0}=
\left|\half+m+i\right|^2 \frac{\pi}{2^{2+2n_s}}\Gamma ^2\left( 1-n_s\right)
 z_{eq}^{-1} g_1^2 z_S^{-2m-5}\left( \frac{g_S}{g_1}\right) ^{2l}
\frac{\sqrt{M^2+\omega^2}}{\omega_s}
\left( \frac {\omega}{\omega_s}\right)^{3+2n_s}
\left( \frac{M}{M_s}\right)^{m-1/2}_{.} 
\label{pirdng0}
\end{equation}

{\em (b) The case $n_1<0$.}\\
In the case $n_1<0$, the approximate solution for $\eta \laq\eta _{re}$ is
given in eq.(\ref{pnle0})  
and for $\eta \gaq \eta _{re}$ it is given in eq.(\ref{pmassRD}).
We connect the solutions at $\eta=\eta_{re}$ by matching them and 
their first derivatives,  
$\Pi_k^{out}(\eta _{re})=\Pi_k^{massive}(\eta _{re})$ and
${\Pi_k^{out}}(\eta _{re})'={\Pi_k^{massive}}(\eta _{re})'$.
Using $k_{re}/k_s=\left( Ma_1/k_s\right)^{1/2}z_S$,  
and since we assume as in
\cite{peak} $k_1/a_1=M_s$, we obtain
\begin{equation}
\left| \beta _k\right|=\left| -\half-m+i\right| 
\frac{\sqrt{\pi}\ 2^{4n_1-n_s-4} } { n_1  (1-2n_1)}
\left| \Gamma \left( -n_s^\pi \right)\right|  
 z_S^{-5/2-m} \left( \frac{g_S}{g_1}\right)^l
\left( \frac {\omega}{\omega_s}\right) ^{2+n_s}
\left( \frac{M}{M_s}\right)^{n_1/2-1}_{.} 
\end{equation}
Using eq.(\ref{modef}), and the definitions of $n_S^\pi$ and
$n_1$, we obtain the final result
\begin{eqnarray}
\left[ \frac{d\Omega _\pi }{d\ln \omega}\right]_{n_1<0} &=&
\left|-\half-m+i\right|^2 \pi\ \frac{ 2^{-6-2n_s-8m}}{ m^2 (\half-m)^2}
 \Gamma ^2\left( 1-n_s\right)  \times \nonumber \\ &&
 z_{eq}^{-1} g_1^2 z_S^{-9-2m}
\left( \frac{g_S}{g_1}\right)^{2l}\frac{\sqrt{M^2+\omega^2}}{\omega_s}
\left( \frac {\omega}{\omega_s}\right) ^{7+2n_s}
\left( \frac{M}{M_s}\right)^{-m-3/2}_{.} 
\label{pirdnl0}
\end{eqnarray}

\section{Spectra of produced particles}

In this section we collect the different contributions that we calculated in
the previous section and put them together into complete spectra for massless
and massive particles. We use this opportunity to recall the parameters that
appear in the expressions for the spectra. The spectra we obtain depend on two
parameters, the total redshift during the string phase $z_S=a_1/a_S$, and the
ratio of string coupling at the beginning and end of string phase $g_S/g_1$.
The coupling $g_1$ is considered a known number, the value of the string
coupling today, and so is the string mass $M_s$. The frequency $\omega_s$ is
expressed as $\omega_s=w_1/z_S$ where $\omega_1$ is taken to be  known
\cite{peak}. Also appearing are the redshift since matter-radiation equality
$z_{eq}$. We recall also the definition,
$n_s=\left[\sqrt{3}(l+m)+1-m\right]/2$. All spectra are valid for 
$\omega\leq\omega_s$, unless otherwise stated. We present the resulting
spectra in their original form, but also in a more symmetric form in the
spirit of \cite{sbarduality}, to help in the possible uncovering and
understanding of some underlying symmetry.

\subsection{The spectrum of massless particles}

{\em (a) The case $n_1>1$ ($m<-\half$).}\\
In this case 
$\frac{d\Omega }{d\ln \omega}=
\left[ \frac{d\Omega _\psi }{d\ln \omega}\right]_{n_1>1}+
\left[ \frac{d\Omega _\pi }{d\ln \omega }\right] _{n_1>0}$. 
The relevant expressions appear in eq.(\ref{psing1}) and (\ref{ping0}).
\begin{eqnarray}
\frac{d\Omega }{d\ln \omega}
&=&\frac {2^{2n_s-2m-1}}{(\quarter-m^2)^2} 
\frac{\Gamma ^2\left( n_s\right) }{\Gamma^2\left(m-\half\right) }
z_{eq}^{-1} g_1^2 z_S^{-6} 
\left( \frac{ g_S}{g_1}\right) ^{-2l}
\left( \frac {\omega}{\omega_s}\right)^{7+2m-2n_s}
\nonumber \\
&+&2^{1-2m-2n_s}
\frac{\Gamma^2\left(1 -n_s \right) }{\Gamma ^2\left( m+\half \right) } 
z_{eq}^{-1} g_1^2 z_S^{-4m-4} \left( \frac{g_S}{g_1}\right)^{2l}
\left( \frac {\omega}{\omega_s}\right) ^{3+2m+2n_s}_{.}
\label{mlessI}
\end{eqnarray}

The leading contribution is determined by the ratio of the two terms which is
approximately $z_S^{2-4m} (\frac{g_S}{g_1})^{4l}$,  if $\frac{g_S}{g_1}\laq 1$
and $z_S\gg1$ then the second term will dominate. The frequency dependence of
the second (and leading) term in eq.(\ref{mlessI}) is determined by the index
$n=3+2m+2n_s=4+m+\sqrt{3}(l+m)$. So, for example, for a perturbative dilaton
dependence $l=-1$, $n=4-\sqrt{3}+(1+\sqrt{3})m$. Therefore, all spectra with
$l=-1$ and $m\le-1$ will be decreasing spectra. In particular, the case $l=-1$
and $m=-1$ corresponding to the usual antisymmetric tensor will have
$n=3-2\sqrt{3}\simeq -0.46$. More negative values of $m$ will 
give more sharply decreasing spectra.

We may present eq.(\ref{mlessI}) in a more symmetric form suggested in
\cite{sbarduality}. We do this to
allow for a future study to  reveal  underlying symmetry.
\begin{eqnarray}
\frac{d\Omega }{d\ln \omega}
&=& {\cal N}_{l,m}^I\ 
z_{eq}^{-1} g_1^2 z_S^{-2m-5}\left( \frac {\omega}{\omega_s}\right)^{5+2m}
\nonumber \\ &\times&
\Biggl\{{\cal A}_{l,m}^I\ z_S^{2m-1}
\left( \frac{g_S}{g_1}\right)^{-2l}
\left( \frac {\omega}{\omega_s}\right)^{2-2n_s}
\! +\!\! {{\cal A}_{l,m}^I}^{-1}\ 
z_S^{-2m+1}\left( \frac{g_S}{g_1}\right) ^{2l}
\left( \frac {\omega}{\omega_s}\right) ^{-2+2n_s}\Biggr\},
\end{eqnarray}
where 
${\cal N}_{l,m}^{II}$ and ${\cal A}_{l,m}^{II}$ can be read off
eq.(\ref{mlessI}).

{\em (b) The case $1>n_1>0$ ($\half>m>-\half$).}\\
In this case
$\frac{d\Omega }{d\ln \omega}=
\left[ \frac{d\Omega _\psi }{d\ln \omega}\right]_{n_1<1}+
\left[ \frac{d\Omega _\pi }{d\ln \omega }\right]_{n_1>0}$.
The relevant expressions are given in eq.(\ref{psinl1}) and (\ref{ping0}).
\begin{eqnarray}
\frac{d\Omega }{d\ln \omega}&=&
2^{2n_s+2m-1}
\frac{\Gamma^2\left( n_s\right) }{\Gamma ^2\left(\half-m\right) } 
z_{eq}^{-1} g_1^2 z_S^{4m-4} \left( \frac{g_S}{g_1}\right)^{-2l}
\left( \frac {\omega}{\omega_s}\right)^{5-2m-2n_s}
\nonumber \\
&+&2^{1-2m-2n_s}
\frac{\Gamma^2\left(1 -n_s \right) }{\Gamma ^2\left( m+\half \right) } 
z_{eq}^{-1} g_1^2 z_S^{-4m-4} \left( \frac{g_S}{g_1}\right)
^{2l}\left( \frac {\omega}{\omega_s}\right) ^{3+2m+2n_s}_{.}
\label{mlessII}
\end{eqnarray}
The leading contribution is determined by the ratio of the two terms which is
approximately $z_S^{-8m} (\frac{g_S}{g_1})^{4l}$, 
if $\frac{g_S}{g_1}\laq 1$ and $z_S\gg1$ then the dominant term is determined
by the sign of $m$. The case $m=0$, $l=-1$ corresponds to photons, studied in
\cite{photons}. In that case, the second term dominates because of the factor
$(g_1/g_S)^4$. It is the only term
computed in \cite{photons}. The index determining the
frequency dependence is $n=4-\sqrt{3}$. It is comforting to observe that our
method of calculation, not relying on specific background evolution during the
string phase gives the same results (where available) as a calculation relying
on some specific background evolution.

We may represent eq.(\ref{mlessII}) in a more symmetric form. We do this to
allow for a future study to  reveal  underlying symmetry.
\begin{eqnarray}
\frac{d\Omega }{d\ln \omega}
&=& {\cal N}_{l,m}^{II}\ 
z_{eq}^{-1}z_S^{-4}g_1^2\left( \frac {\omega}{\omega_s}\right)^4 \nonumber \\
&\times&\Biggl\{{\cal A}_{l,m}^{II}
\left( \frac{g_S}{g_1}\right)^{-2l}
\left( \frac{\omega}{\omega_s}\right)^{1-2m-2n_s}+ 
{{\cal A}_{l,m}^{II}}^{-1}\left( \frac{g_S}{g_1}\right)^{2l}
\left( \frac{\omega}{\omega_s}\right) ^{-1+2m+2n_s}\Biggr\}, 
\end{eqnarray}
where 
${\cal N}_{l,m}^{II}$ and ${\cal A}_{l,m}^{II}$ can be read off
eq.(\ref{mlessII}).

{\em (c) The case $n_1<0$ ($m>\half$).}\\
In this case
$\frac{d\Omega }{d\ln \omega}=
\left[ \frac{d\Omega _\psi }{d\ln \omega}\right]_{n_1<1}+
\left[ \frac{d\Omega _\pi }{d\ln \omega }\right] _{n_1<0}$.
The relevant expressions are given in eq.(\ref{psinl1}) and (\ref{pinl0}).
\begin{eqnarray}
\frac{d\Omega }{d\ln \omega}&=& 2^{2n_s+2m-1}
\frac{\Gamma^2\left( n_s\right) }{\Gamma ^2\left(\half-m\right) } 
z_{eq}^{-1} g_1^2 z_S^{4m-4} \left( \frac{g_S}{g_1}\right)^{-2l}
\left( \frac {\omega}{\omega_s}\right) ^{5-2m-2n_s}
\nonumber \\
&+& \frac{2^{-6m -2n_s-5}}{m^2 (m-\half)^2}
\frac{\Gamma ^2\left( 1-n_s \right)}{\Gamma ^2\left( -\half-m \right) }
z_{eq}^{-1} g_1^2 z_S^{-6} \left( \frac{g_S}{g_1}\right)^{2l}
\left( \frac {\omega}{\omega_s}\right) ^{5-2m+2n_s}_{.}
\label{mlessIII}
\end{eqnarray}

The leading contribution is determined by the ratio of the two terms which is
approximately $z_S^{-4m-2} (\frac{g_S}{g_1})^{4l}$,  if $\frac{g_S}{g_1}\laq
1$ and $z_S\gg1$ then the first term will dominate. The frequency dependence
of the first (and leading) term in eq.(\ref{mlessIII}) is determined by the
index $n=5-2m-2n_s=4-m-\sqrt{3}(l+m)$. So, for example, for a perturbative
dilaton dependence $l=-1$, $n=4+\sqrt{3}-(1+\sqrt{3})m$. Therefore, all
spectra with $l=-1$ and $m\le 1$ will be increasing spectra. In particular,
the case $l=-1$ and $m=1$ corresponding to gravitons and dilatons will have
$n=3$ as in \cite{pert,bggv} (we have used an approximation in which the
logarithmic factors are missing). Again, we observe that our method
of calculation, not relying on specific background evolution during the string
phase gives the same results (where available) as a calculation relying on
some specific background evolution.  

More negative values of $m$ will correspondingly give  more sharply increasing
spectra. The spectrum of the model independent axion,  for which $l=1$, $m=1$,
will have an index $n=4-2\sqrt{3}\simeq -0.46$. Note that this is exactly the
index of the leading contribution to the spectrum of the antisymmetric tensor
(a fact known to the authors of \cite{bmvu}). In this class of generically
increasing spectra the axion is standing out with it's decreasing spectrum.

We may represent eq.(\ref{mlessIII}) in a more symmetric form. We do this to
allow for a future study to  reveal  underlying symmetry.
\begin{eqnarray}
\frac{d\Omega }{d\ln \omega}
&=& {\cal N}_{l,m}^{III}\ 
z_{eq}^{-1}z_S^{2m-5}g_1^2\left( \frac {\omega}{\omega_s}\right)^{5-2m}
\nonumber \\
&\times& \Biggl\{{\cal A}_{l,m}^{III}\
z_S^{2m+1}\left( \frac{g_S}{g_1}\right) ^{-2l}\left(
\frac {\omega}{\omega_s}\right) ^{-2n_s}
+ {{\cal A}_{l,m}^{III}}^{-1}
z_S^{-2m-1}\left( \frac{g_S}{g_1}\right) ^{2l}\left(
\frac {\omega}{\omega_s}\right) ^{2n_s}\Biggr\},
\end{eqnarray}
where 
${\cal N}_{l,m}^{III}$ and ${\cal A}_{l,m}^{III}$ can be read off
eq.(\ref{mlessIII}).

\subsection{The spectrum of massive particles}

\subsubsection{The discontinuity in the spectrum of the massive particles}

The spectrum we computed in section (3.C) was that of  massive particles that
enter the horizon as massive particles, so at $\eta_{re}$, 
$k/a(\eta_{re})<M$. The spectrum  of massive particles computed in section
(3.B), is valid for massive particles if at $\eta_{re}$, $k/a(\eta_{re})>M$.
In the approximation scheme that we are using, there will be a discontinuity
in the slope of the spectrum of massive particles, at a wavenumber $k_m$,
where $k_m$ is defined by the condition $k_m/a(\eta_m)=M$. The moment $\eta_m$
is the (conformal) time where the approximation  $k/a(\eta_m)>M$ at reentry is
no longer correct.  To compute $k_m$ we switch momentarily to cosmic time.
First, since $t_m$ is during RD (see below), it can be computed from
$M=H(t_m)= H_1 t_1/t_m$.  Since $a(t_m)$ depends on whether $t_m$ is bigger or
smaller then $t_{eq},$ we get that if $t_m<t_{eq}$  (i.e.
$M>H(t_{eq})=10^{-27}ev$), all the particles will get into the horizon in the
RD period. Therefore $a(t_m)=a_1(t_m/t_1)^{1/2}=a_1(H_1/M)^{1/2}$ and since
$H_1=k_1/a_1$ we obtain $a(t_m)=(k_1 a_1/M)^{1/2}$, so 
$
k_m=\left( M a_1 k_1\right)^{1/2}_{.} 
$
Because $ \frac{\omega_m}{\omega_1}=
\frac{k_m}{k_1} $ and $k_1/a_1=M_s$, $ \frac{k_m}{k_1}=\frac{\left(
Ma_1k_1\right) ^{1/2} }{M_s a_1}. $ Therefore \begin{equation}
\frac{k_m}{k_1}=\left( \frac M{M_s}\right)^{1/2}_{.} \end{equation} The
discontinuity of the slope will be at a frequency $\omega_m=\omega_1\left(
\frac {M}{M_s}\right)^{1/2}$. At that frequency 
\begin{equation}
\frac{\omega_m}{\omega_s}=z_S \left( \frac {M}{M_s}\right)^{1/2}_{.}
\label{omegam}
\end{equation}

If $\omega_s>M$, there will be an additional change in the slope at
$\omega=M$, above which $\frac{\sqrt{M^2+\omega^2}}{\omega_s}\simeq
\frac{\omega}{\omega_s}$, and the spectrum of massive particles and massless
particles become approximately the same.

\subsubsection{Spectrum of massive particles}

{\em (a) The case $n_1>1$ $\left( m<-\half\right).$}\\
In this case
$\frac{d\Omega }{d\ln \omega}=
\left[ \frac{d\Omega _\psi }{d\ln \omega}\right]_{n_1>1}+
\left[ \frac{d\Omega _\pi }{d\ln \omega}\right] _{n_1>0}$.

For $\omega_m<\omega<\omega_s$ the relevant expressions are given in
eq.(\ref{psing1mass}) and  (\ref{ping0mass}).
\begin{eqnarray}
\frac{d\Omega }{d\ln \omega}&=&
\frac {2^{2n_s-2m-1}}{(\quarter-m^2)^2} 
\frac{\Gamma ^2\left( n_s\right) }{\Gamma^2\left(m-\half\right) }
z_{eq}^{-1} g_1^2 z_S^{-6}\left( \frac{g_S}{g_1}\right)^{-2l}
\left( \frac {\omega}{\omega_s}\right)^{6+2m-2n_s}
\frac{\sqrt{M^2+\omega^2}}{\omega_s}
\nonumber \\ &+&
2^{1-2m-2n_s}
\frac{\Gamma^2\left(1 -n_s \right) }{\Gamma ^2\left( m+\half \right) } 
z_{eq}^{-1} g_1^2 z_S^{-4m-4} \left( \frac{g_S}{g_1}\right) ^{2l}
\left( \frac{\omega}{\omega_s}\right)^{2+2m+2n_s}
\frac{\sqrt{M^2+\omega^2}}{\omega_s}.
\label{massspecIa}
\end{eqnarray}

For $\omega<\omega_m$,
the relevant expressions are given in eq.(\ref{mrdng1}) and 
(\ref{pirdng0}).
\begin{eqnarray}
\frac{d\Omega }{d\ln \omega}&=& 
 \frac{\left|3m+i\right| ^2}{\pi }
\frac{2^{2n_s-6} }{ (\quarter-m^2)^2} \Gamma ^2\left( n_s\right)
 z_{eq}^{-1} g_1^2 z_S^{2m-7}\left(\frac{g_S}{g_1}\right)^{-2l}
\left( \frac {\omega}{\omega_s}\right)^{7-2n_s} 
\frac{\sqrt{M^2+\omega^2}}{\omega_s}
\left( \frac{M}{M_s}\right)^{m-1/2}
\nonumber \\  && \hspace{-18pt} +
\left|\half+m+i\right|^2 \frac{\pi}{2^{2+2n_s}}\Gamma^2\left(1-n_s\right)
 z_{eq}^{-1} g_1^2 z_S^{-2m-5}\left( \frac{g_S}{g_1}\right)^{2l}
\left( \frac {\omega}{\omega_s}\right)^{3+2n_s} 
\frac{\sqrt{M^2+\omega^2}}{\omega_s}
\left(\frac{M}{M_s}\right)^{m-1/2}_{.} 
\label{massspecI}
\end{eqnarray}
At $\omega=\omega_m$ we may check, using eq.(\ref{omegam}) that the spectrum
is continuous, up to numerical factors of order 1.
The spectral index, however, jumps at $\omega_m$.

If $\omega_s>M$, then for the range $M<\omega<\omega_s$ the spectrum is
approximately the same as the massless spectrum, eq.(\ref{mlessI}).

As in the massless case we may determine the leading contribution to the
spectrum. It is determined by the ratio of the terms which is
approximately $z_S^{2-4m} (\frac{g_S}{g_1})^{4l}$, 
if $\frac{g_S}{g_1}\laq 1$ and $z_S\gg1$ then the second term will dominate.
For example, for $l=-1$ and  $m=-1$ the spectral index will be
$n=2-2\sqrt{3}\simeq -1.46$ for $w>\omega_m$ and 
$n=5-2\sqrt{3}\simeq +1.53$ for $w<\omega_m$. If $\omega_s>M$, then for
$M<\omega<\omega_s$ the index is $n=3-2\sqrt{3}\simeq -0.46$.

Eqs.(\ref{massspecIa}), (\ref{massspecI})  may be presented in a more
symmetric, perhaps related to some  symmetry of the underlying physics. We
present here only the symmetric form of eq.(\ref{massspecI}).
\begin{eqnarray}
\frac{d\Omega }{d\ln \omega}&=& {\cal M}_{l,m}^{I}
 z_{eq}^{-1} g_1^2 z_S^{-6}\frac{\sqrt{M^2+\omega^2}}{\omega_s}
 \left( \frac {\omega}{\omega_s}\right)
^5\left( \frac M{M_s}\right) ^{m-\half}
\nonumber \\ &\times& 
\Biggl\{ {\cal B}_{l,m}^{I}
\left( \frac{g_S}{g_1}\right) ^{-2l}z_S^{2m-1}
\left( \frac{\omega}{\omega_s}\right) ^{2-2n_s}+ {{\cal B}_{l,m}^{I}}^{-1}
\left( \frac{g_S}{g_1}\right) ^{2l}z_S^{-2m+1}
\left(\frac {\omega}{\omega_s}\right) ^{-2+2n_s}\Biggr\}, 
\end{eqnarray}
where ${\cal M}_{l,m}^{I}$ and ${\cal B}_{l,m}^{I}$ are numerical coefficients
of order one which could be read off eq.(\ref{massspecI}).

{\em (b) The case $1>n_1>0$ $\left( -\half<m<\half\right).$}\\
In this case
$
\frac{d\Omega }{d\ln \omega}=
\left[ \frac{d\Omega _\psi }{d\ln \omega}\right]_{n_1<1}
+\left[ \frac{d\Omega _\pi }{d\ln \omega}\right] _{n_1>0}$.

For $\omega_m<\omega<\omega_s$ the relevant expressions are given in
eq.(\ref{psinl1mass}) and  (\ref{ping0mass}).
\begin{eqnarray}
\frac{d\Omega }{d\ln \omega}&=& 
2^{2n_s+2m-1}
\frac{\Gamma^2\left( n_s\right) }{\Gamma ^2\left(\half-m\right) } 
z_{eq}^{-1} g_1^2 z_S^{4m-4}\left(\frac{g_S}{g_1}\right)^{-2l}
\left( \frac{\omega}{\omega_s}\right)^{4-2m-2n_s}
\frac{\sqrt{M^2+\omega^2}}{\omega_s}
\nonumber \\ &+&
2^{1-2m-2n_s}
\frac{\Gamma^2\left(1 -n_s \right) }{\Gamma ^2\left( m+\half \right) } 
z_{eq}^{-1} g_1^2 z_S^{-4m-4} \left( \frac{g_S}{g_1}\right) ^{2l}
\left( \frac{\omega}{\omega_s}\right)^{2+2m+2n_s}
\frac{\sqrt{M^2+\omega^2}}{\omega_s}.
\label{massspecIIa}
\end{eqnarray}
For $\omega<\omega_m$,
the relevant expressions are given in eq.(\ref{mrdnl1}) and 
(\ref{pirdng0}).
\begin{eqnarray}
\frac{d\Omega }{d\ln \omega}&=& \frac{\left|\half+m+i\right| ^2}{\pi }
2^{2n_s-4}\Gamma ^2\left( n_s\right)
z_{eq}^{-1} g_1^2 z_S^{2m-3}\left( \frac{g_S}{g_1}\right)^{-2l}
\frac{\sqrt{M^2+\omega^2}}{\omega_s}\left( \frac {\omega}{\omega_s}\right)^{3-2n_s}
\left( \frac{M}{M_s}\right)^{1/2-m} 
\nonumber \\ && \hspace{-18pt} +
\left|\half+m+i\right|^2 \frac{\pi}{2^{2+2n_s}}\Gamma ^2\left( 1-n_s\right)
z_{eq}^{-1} g_1^2 z_S^{-2m-5}\left( \frac{g_S}{g_1}\right) ^{2l}
\frac{\sqrt{M^2+\omega^2}}{\omega_s}
\left( \frac {\omega}{\omega_s}\right)^{3+2n_s}
\left( \frac{M}{M_s}\right)^{m-1/2}_{.} 
\label{massspecII}
\end{eqnarray}
At $\omega=\omega_m$ we may check, using eq.(\ref{omegam}) that the spectrum
is continuous, up to numerical factors of order 1.

If $\omega_s>M$, then for the range $M<\omega<\omega_s$ the spectrum is
approximately the same as the massless spectrum, eq.(\ref{mlessII}). 

The determination of the leading contribution is more complicated now, because
the ratio of the two terms in each range is different and involves also
the ratio $\frac{M}{M_s}$. We postpone such analysis to a more detailed
study of specific cases.

Eqs.(\ref{massspecIIa}), (\ref{massspecII})  may be presented in a more
symmetric, perhaps related to some  symmetry of the underlying physics. We
present here only the symmetric form of eq.(\ref{massspecII}).
\begin{eqnarray}
&&\frac{d\Omega }{d\ln \omega}={\cal M}_{l,m}^{II}
 z_{eq}^{-1}z_S^{-4}g_1^2\frac{\sqrt{M^2+\omega^2}}{\omega_s}
 \left( \frac {\omega}{\omega_s}\right)^3
 \nonumber \\ && \hspace{-4pt}  \times   
\Biggl\{ {{\cal B}_{l,m}^{II}}
z_S^{2m+1}\left( \frac{g_S}{g_1}\right) ^{-2l}
\left( \frac{\omega}{\omega_s}\right)^{-2n_s}
\left( \frac M{M_s}\right) ^{m-\half}\!\!+ {{\cal B}_{l,m}^{II}}^{-1}
z_S^{-1-2m}\left( \frac{g_S}{g_1}\right)^{2l}
\left( \frac {\omega}{\omega_s}\right)^{2n_s}
\left( \frac{M}{M_s}\right) ^{\half-m}\Biggr\},
\end{eqnarray}
where ${\cal M}_{l,m}^{II}$ and ${\cal B}_{l,m}^{II}$ are numerical
 coefficients of order one which could be read off eq.(\ref{massspecII}).

{\em (c) The case $n_1<0$ $\left( m>\half\right).$}\\ 
In this case
$
\frac{d\Omega }{d\ln \omega}=
\left[ \frac{d\Omega _\psi }{d\ln \omega}\right]_{n_1<1}+
\left[ \frac{d\Omega _\pi }{d\ln \omega}\right]_{n_1<0}.
$

For $\omega_m<\omega<\omega_s$ the relevant expressions are given in
eq.(\ref{psinl1mass}) and  (\ref{pinl0mass}).
\begin{eqnarray}
\frac{d\Omega }{d\ln \omega}&=&
2^{2n_s+2m-1}
\frac{\Gamma^2\left( n_s\right) }{\Gamma ^2\left(\half-m\right) } 
z_{eq}^{-1} g_1^2 z_S^{4m-4}\left(\frac{g_S}{g_1}\right)^{-2l}
\left( \frac{\omega}{\omega_s}\right)^{4-2m-2n_s}
\frac{\sqrt{M^2+\omega^2}}{\omega_s}
\nonumber \\ &+&
\frac{2^{-6m -2n_s-5}}{m^2(m-\half)^2}
\frac{\Gamma ^2\left( 1-n_s \right)}{\Gamma ^2\left( -\half -m \right) }
z_{eq}^{-1} g_1^2 z_S^{-6} \left( \frac{g_S}{g_1}\right) ^{2l}
\left( \frac {\omega}{\omega_s}\right)^{4-2m+2n_s}
\frac{\sqrt{M^2+\omega^2}}{\omega_s}.
\label{massspecIIIa}
\end{eqnarray}
For $\omega<\omega_m$,
the relevant expressions are given in eq.(\ref{mrdnl1}) and 
(\ref{pirdnl0}).
\begin{eqnarray}
&&\left[ \frac{d\Omega _\psi }{d\ln \omega}\right]=
\frac{\left|\half+m+i\right| ^2}{\pi }
2^{2n_s-4}\Gamma ^2\left( n_s\right)
z_{eq}^{-1} g_1^2 z_S^{2m-3}\left( \frac{g_S}{g_1}\right)^{-2l}
\frac{\sqrt{M^2+\omega^2}}{\omega_s}
\left( \frac {\omega}{\omega_s}\right)^{3-2n_s}
\left( \frac{M}{M_s}\right)^{1/2-m} + 
\nonumber \\ &&
\left|\half\!+m\!-i\right|^2 \pi\ \frac{ 2^{-6-2n_s-8m}}{ m^2(\half\!-\!m)^2}
\Gamma ^2\left(1\!-\!n_s\right)  
 z_{eq}^{-1} g_1^2 z_S^{-9-2m}
\left( \frac{g_S}{g_1}\right)^{2l}\frac{\sqrt{M^2+\omega^2}}{\omega_s}
\left( \frac {\omega}{\omega_s}\right)^{7+2n_s}
\left( \frac{M}{M_s}\right)^{-m-3/2}_{.} 
\label{massspecIII}
\end{eqnarray}
At $\omega=\omega_m$ we may check, using eq.(\ref{omegam}) that the spectrum
is continuous, up to numerical factors of order 1.

If $\omega_s>M$, then for the range $M<\omega<\omega_s$ the spectrum is
approximately the same as the massless spectrum, eq.(\ref{mlessIII}).

The determination of the leading contribution is more complicated now, because
the ratio of the two terms in each range is different and involves also
the ratio $\frac{M}{M_s}$. We postpone such analysis to a more detailed
study of specific cases.

Eqs.(\ref{massspecIIIa}), (\ref{massspecIII})  may be presented in a more
symmetric, perhaps related to some  symmetry of the underlying physics. We
present here only the symmetric form of eq.(\ref{massspecIII}).
\begin{eqnarray}
&& \frac{d\Omega }{d\ln \omega}= {{\cal M}_{l,m}^{III}}
z_{eq}^{-1}z_S^{-6}g_1^2
\frac{\sqrt{M^2+\omega^2}}{\omega_s}\left( \frac {\omega}{\omega_s}\right)^5
\left( \frac {M}{M_s}\right) ^{-\half-m} 
\nonumber \\ && \hspace{-24pt} \times 
\Biggl\{ {{\cal B}_{l,m}^{III}}
z_S^{2m+3}\left( \frac{g_S}{g_1}\right) ^{-2l}
\left( \frac{\omega}{\omega_s}\right) ^{-2-2n_s}
\left( \frac {M}{M_s}\right) +{{\cal B}_{l,m}^{III}}^{-1}
z_S^{-3-2m}\left( \frac{g_S}{g_1}\right) ^{2l}
\left( \frac {\omega}{\omega_s}\right) ^{2+2n_s}
\left( \frac{M}{M_s}\right)^{-1}\Biggr\},
\end{eqnarray}
where ${\cal M}_{l,m}^{III}$ and ${\cal B}_{l,m}^{III}$ are numerical 
coefficients of order one which could be read off eq.(\ref{massspecIII}).

The spectra of massless and massive particles can be translated into each
other by using the relations  $M/M_s=\left( k_{re}/k_s\right)^2 z_S^{-2}$ for
the massive case and  $\left( k_{re}/k_s\right)^2 z_S^{-2}= \left(
\omega/\omega_s\right) ^2z_S^{-2}$ for the massless case. The substitutions 
$\left( \omega/\omega_s\right) ^2z_S^{-2} \leftrightarrow M/M_s$ in addition
to  the substitution accounting for the difference in the energy
$\sqrt{M^2+\omega^2}/\omega_s\leftrightarrow \omega/\omega_s$ translate the
results from the massive case into those of the massless case and vice versa.

\section{Conclusions}

Spectra of produced particles in string cosmology models come in different
shapes, as summarized in section 4. Even though the background
curvature is increasing during the inflationary dilaton-driven phase, spectra
are not necessarily increasing. This peculiarity comes about because different
types of particles couple differently to the background curvature and dilaton,
and in addition, spectra of massless and massive particles are different.

During the string phase, the use of the lowest order equations for the
background or for perturbations is questionable,  and therefore it might have
been possible to doubt the validity and accuracy of the calculations of
particle production. We have argued that to estimate particle production
during the dilaton-driven phase it is not necessary to know the details of the
evolution during the string phase. If the principle of causality is used,
expressed in practical terms by the ``freezing" of the perturbation amplitude
and it's conjugate momentum, it is possible to do reliable calculations. Our
results provide several explicit checks of this principle, by  comparison with
other calculations, and therefore give further credibility to existing
spectral calculations.

We have not used any explicit symmetry considerations in our calculations.
However, it is clear that spectral indices and perhaps other properties of the
spectra could be analized using an underlying symmetry as in
\cite{sbarduality}.

Among the spectra we found there are some with weak dependence on frequency
and some decreasing spectra which contain substantial power at large
wavelength. Because these spectra depend differently on the basic parameters
of the models, requiring that they  be compatible with  astrophysical and
cosmological bounds is likely to narrow the allowed range of the parameters.
Their existence also suggests an obvious source for the observed large scale
anisotropy. 

Relic massive particles are an appealing source for cold or hot dark matter.
However, to decide on this issue, further analysis and input are necessary. 
To compute relic abundances of particles we need to know more about
their interactions and also about the late time evolution of the Universe. We
are completing such analysis for axions, and hope to perform similar
analyses for other particles.

As already mentioned several times, there are obvious improvements and
generalizations to our calculations. Straightforward improvements, that cannot
however be performed in a general way, concern the late time background
evolution, and the inclusion of effects of particle decay and interactions.
Among possible generalizations, we view considering other  background
evolutions as particularly interesting. This was partially done in
\cite{axions,bmvu} and should provide information
about the robustness of the spectra.

\acknowledgments 

We would like to thank A. Buonanno, K. Meissner, C. Ungarelli and G. Veneziano
for sharing with us unpublished results and for useful discussions of several
aspects of our calculations. This work is supported in part by the  Israel
Science Foundation administered by the Israel Academy of Sciences and
Humanities.


\begin{references}
\bibitem{comm}
We have ignored a numerical factor in the definition of $\Omega$, which in some
conventions is $8\pi/3$. We ignore this factor throughout.
\end{references}
\end{document}